\documentclass[letterpaper,12pt,notoc]{JHEP3}
\def\pd{\partial}

\usepackage{graphicx}
\usepackage{amsmath}
\usepackage{amssymb}

\preprint{ \hbox{}\hfill arXiv: 1202.1043}

\title{3D N=6 Gauged Supergravity: Admissible Gauge Groups,Vacua and
RG Flows}
\author{Auttakit Chatrabhuti$^{a, \, b}$, Parinya Karndumri$^{a,\, b}$ and Boonpithak Ngamwatthanakul$^a$\\
$^a$String Theory and Supergravity Group, Department
of Physics, Faculty of Science, Chulalongkorn University, 254 Phayathai Road, Pathumwan, Bangkok 10330, Thailand\\
$^b$Thailand Center of Excellence in Physics, CHE, Ministry of Education, Bangkok 10400, Thailand \\
E-mail: \email{auttakit@sc.chula.ac.th},
\email{parinya.ka@hotmail.com},\email{boonpithak@gmail.com}}

\abstract{We study $N=6$ gauged supergravity in three dimensions
with scalar manifolds $\frac{SU(4,k)}{S(U(4)\times U(k))}$ for
$k=1,2,3,4$ in great details. We classify some admissible
non-compact gauge groups which can be consistently gauged and
preserve all supersymmetries. We give the explicit form of the
embedding tensors for these gauge groups as well as study their
scalar potentials on the full scalar manifold for each value of
$k=1,2,3,4$ along with the corresponding vacua. Furthermore, the
potentials for the compact gauge groups, $SO(p)\times SO(6-p)\times
SU(k)\times U(1)$ for $p=3,4,5,6$, identified previously in the
literature are partially studied on a submanifold of the full scalar
manifold. This submanifold is invariant under a certain subgroup of
the corresponding gauge group. We find a number of supersymmetric
AdS vacua in the case of compact gauge groups. We then consider
holographic RG flow solutions in the compact gauge groups
$SO(6)\times SU(4)\times U(1)$ and $SO(4)\times SO(2)\times
SU(4)\times U(1)$ for the $k=4$ case. The solutions involving one
active scalar can be found analytically and describe operator flows
driven by a relevant operator of dimension $\frac{3}{2}$. For
non-compact gauge groups, we find all types of vacua namely AdS,
Minkowski and dS, but there is no possibility of RG flows in the
AdS/CFT sense for all gauge groups considered here.}
\begin{document}
\section{Introduction}
Gauged supergravity is interesting on its own right and has many
applications in string theory. It can give some insight to the study
of flux compactifications, see \cite{gaugeSUGRA_flux} for a review,
as well as many applications in the context of the AdS/CFT
correspondence \cite{maldacena}. In the present work, we will
concentrate on the latter application. This involves classifying the
AdS vacua or AdS critical points of the scalar potential which is in
turn, dual to some CFT's living on the boundary of the $AdS$ space.
We will also consider RG flows between these CFT's. On the gravity
side, the RG flows are described by domain wall solutions connecting
two vacua. We focus our attention on gauged supergravity in three
dimensions that is useful in the study of AdS$_3$/CFT$_2$
correspondence which recently also has some applications in
condensed matter systems \cite{henning_ADSCMT}.
\\
\indent The ungauged supergravity theories in three dimensions with
all possible numbers of supersymmetries, $1\leq N\leq 16$, have been
constructed in \cite{dewit1}, and their gauged version has
subsequently been studied in \cite{dewit}. The gauged supergravity
is described in the form of Chern-Simons gauged supergravity in
which the gauge fields enter the Lagrangian via Chern-Simons term.
All the bosonic degrees of freedom are carried by scalar fields
which, for $N>4$, are encoded in a symmetric space of the form $G/H$
\cite{dewit1} where $G$ is the global symmetry of the theory, and
$H$ is its the maximal compact subgroup. In this sense, the vector
fields with Chern-Simons term will not introduce any new degrees of
freedom. Unlike in higher dimensional analogues, the number of gauge
fields, or equivalently the dimension of the gauge group, is not
fixed. Therefore, there are more possibilities for the choices of
gauge groups.
\\
\indent It is more convenient, particularly with the formulation of
\cite{dewit} in which the classification of admissible gauge groups
can be done in a $G$-covariant way, to classify the possible gauge
groups in the Chern-Simons form rather than in the more conventional
Yang-Mills gauged supergravity. Of course, the latter is related to
some higher dimensional theories via certain dimensional reductions.
Apart from the case of non-semisimple gaugings as discussed in
\cite{csym}, the Chern-Simons and Yang-Mills gauged supergravities
are not equivalent. This fact has a well-known consequence that a
Chern-Simons gauged supergravity with other types of gauge groups,
compact and non-compact, cannot be obtained from any known higher
dimensional framework. Since we will not consider non-semisimple
gauge groups in this work, by gauged supergravity, we always mean
Chern-Simons gauged supergravity.
\\
\indent We are interested in $N=6$ gauged supergravity in three
dimensions which has been studied in the context of superconformal
gaugings in \cite{3D_conformal_gauge}. In this reference, the
conformal gaugings of the $N=6$ theory give rise to superconformal
field theories in three dimensions which can be regarded as theories
on the worldvolume of M2-branes and are relevant in the discussion
of AdS$_4$/CFT$_3$ correspondence. In the present work, we will
study the $N=6$ theory in the context of AdS$_3$/CFT$_2$
correspondence. In this way, we study $N=6$ gauged supergravity as a
supergravity theory rather than its global supersymmetry limit. The
scalar manifold of this theory is of the form
$\frac{SU(4,k)}{S(U(4)\times U(k))}$ where $k$ is the number of
matter multiplets, see $\cite{dewit1}$ for more details. We will
study four cases namely $k=1,2,3,4$.
\\
\indent The gaugings are implemented by mean of the embedding
tensor. This tensor has to satisfy quadratic and linear constraints
in order that the gaugings are consistent and compatible with
supersymmetry. The embedding tensor for compact gauge groups has
been given in \cite{dewit}. In this work, we will identify some
non-compact gauge groups which can be consistently gauged. We then
find the corresponding scalar potentials for each gauge group both
compact and non-compact and study their critical points as well.
Critical points of gauged supergravities in three dimensions with
different numbers of supersymmetries have been obtained in many
places \cite{bs}, \cite{gkn}, \cite{AP}, \cite{AP2}, \cite{Fisch}
and \cite{N16Vacua}.
\\
\indent We will also consider RG flow solutions interpolating
between two AdS$_3$ critical points in the case of compact gauge
groups. According to the AdS/CFT correspondence, these solutions
describe RG flows between two conformal fixed points of a two
dimensional dual field theory. Within the framework of three
dimensional gauged supergravities, this study has extensively been
explored in a number of previous works \cite{bs}, \cite{gkn},
\cite{AP}, \cite{deger}. The structure of the critical points allows
us to find RG flow solutions with only one active scalar that has a
non-trivial dependence on the radial coordinate. This gives rise to
a simple flow equation that can be solved analytically.
\\
\indent The paper is organized as follows. We review the formulation
of three dimensional gauged supergravity in section \ref{3DSUGRA}.
After a general discussion, we specify to the case of $N=6$ theory.
In section \ref{compact}, we study scalar potentials for compact
groups as well as their critical points. In section
\ref{non_compact}, we classify some non-compact gauge groups which
can be consistently gauged. We then find their scalar potentials and
the corresponding critical points. The RG flow solutions for
$SO(6)\times SU(4)\times U(1)$ and $SO(4)\times SO(2)\times
SU(4)\times U(1)$ gauge groups in the $k=4$ case are given in
section \ref{flow}. We also give some detail of the Euler angle
parametrization in the appendix including an explicit example. We
end the paper with some comments and summary of the main results.
\section{Three dimensional gauged supergravity}\label{3DSUGRA}
In this section, we give a brief construction of three dimensional
gauged supergravity using the formulation given in \cite{dewit}. We
begin with some general features and useful formulae which play an
important role in various places of the paper. We finally specify to
the $N=6$ theory in which scalar fields are encoded in the symmetric
space $\frac{SU(4,k)}{S(U(k)\times U(4))}$. We refer the reader to
\cite{dewit} for the full detail of the construction.
\subsection{General construction}
It has been shown in \cite{dewit1} that matter coupled supergravity
in three dimensions is in the form of a non-linear sigma model
coupled to pure supergravity. The target space manifold of scalars,
for $N>4$, is a symmetric space of the form $G/H$ with global and
local symmetry groups $G$ and $H$, respectively. The group $H$
contains the R-symmetry $SO(N)$ and is of the form $H=SO(N)\times
H'$. The $G$ generators $t^{\mathcal{M}}$ can be decomposed
accordingly into $(X^{IJ},X^\alpha, Y^A)$ where $X^{IJ}$ and
$X^\alpha$ generate $SO(N)$ and $H'$, respectively, and $Y^A$ are
non-compact or coset generators. These generators satisfy the $G$
algebra
\begin{eqnarray}
\left[X^{IJ},X^{KL}\right]&=&-4\delta^{[I[K}X^{L]J]}, \qquad
\left[X^{IJ},Y^A\right]=-\frac{1}{2}f^{IJ,AB}Y_B, \nonumber \\
\left[X^\alpha,X^\beta\right]&=&f^{\alpha
\beta}_{\phantom{as}\gamma}X^\gamma,\qquad
\left[X^\alpha,Y^A\right]=h^{\alpha
\phantom{a}A}_{\phantom{a}B}Y^B, \nonumber \\
\left[Y^{A},Y^{B}\right]&=&\frac{1}{4}f^{AB}_{IJ}X^{IJ}+\frac{1}{8}C_{\alpha\beta}h^{\beta
AB}X^\alpha \label{Galgebra}
\end{eqnarray}
where $C_{\alpha\beta}$ and $h^{\alpha \phantom{a}A}_{\phantom{a}B}$
are an $H'$ invariant tensor and anti-symmetric tensors defined in
\cite{dewit1}.
\\
\indent In parametrizing the symmetric space $G/H$, we introduce a
coset representative $L$ which transforms under $G$ and $H$ as
$L\rightarrow gLh$ where $g\in G$ and $h\in H$. Some useful formulae
for a coset space are
\begin{equation}
L^{-1} \partial_i L= \frac{1}{2}Q^{IJ}_i X^{IJ}+Q^\alpha_i
X^{\alpha}+e^A_i Y^A,\label{cosetFormula1}
\end{equation}
and
\begin{equation}
L^{-1}t^\mathcal{M}L=\frac{1}{2}\mathcal{V}^{\mathcal{M}IJ}X^{IJ}+\mathcal{V}^\mathcal{M}_{\phantom{a}\alpha}X^\alpha+
\mathcal{V}^\mathcal{M}_{\phantom{a}A}Y^A\, .\label{cosetFormula}
\end{equation}
In the above formulae, $Q^{IJ}_i$ and $Q^\alpha_i$ are composite
connections for $SO(N)$ and $H'$, respectively. $I, J,\ldots
=1,\ldots, N$ denote the R-symmetry indices, and $\alpha,
\beta,\ldots =1,\ldots, \textrm{dim} H'$ are $H'$ adjoint indices.
The vielbein $e^A_i$ can be used to construct the metric on the
scalar manifold via
\begin{equation}
g_{ij}=e^A_ie^B_j\delta_{AB},\qquad i, j, A, B =1,\ldots,
\textrm{dim}\, (G/H)
\end{equation}
and together with its inverse can be used to interchange between
curve and flat target space indices. The $\mathcal{V}$'s will be
used to define the T-tensors which in turn are needed in order to
find the scalar potential. We will come back to this later.
\\
\indent Following \cite{dewit}, gaugings are described by
introducing an embedding tensor $\Theta_{\mathcal{M}\mathcal{N}}$.
This tensor is symmetric in its indices and gauge invariant. It acts
as a projector on the symmetry group $G$ to the gauge group
$G_0\subset G$. The requirement that the gauge generators given by
\begin{equation}
J_{\mathcal{M}}=\Theta_{\mathcal{MN}}t^{\mathcal{N}}
\end{equation}
form an algebra imposes the so-called quadratic constraint on the
embedding tensor
\begin{equation}
\Theta_{\mathcal{PL}}f^{\mathcal{KL}}_{\phantom{asds}\mathcal{(M}}\Theta_{\mathcal{N)K}}=0\,
.\label{theta_quadratic}
\end{equation}
Furthermore, there is a projection constraint which is a consequence
of the requirement that a given gauging is consistent with
supersymmetry. In general, this constraint is imposed at the level
of the T-tensors defined by
\begin{equation}
T_{\mathcal{A}\mathcal{B}}=\mathcal{V}^{\mathcal{M}}_{\phantom{a}\mathcal{A}}\Theta_{\mathcal{M}\mathcal{N}}
\mathcal{V}^{\mathcal{N}}_{\phantom{a}\mathcal{B}}\, .
\end{equation}
The projection constraint, acting only on the $T^{IJKL}$ component,
can be written as
\begin{equation}
\mathbb{P}_{\boxplus}T^{IJ,KL}=0\label{T_projection}
\end{equation}
where $\boxplus$ denotes the representation $\boxplus$ of $SO(N)$.
One important result from \cite{dewit} is that for symmetric target
spaces, the projection constraint \eqref{T_projection} can be
uplifted to the condition on the embedding tensor
\begin{equation}
\mathbb{P}_{R_0}\Theta_{\mathcal{MN}}=0\, .\label{theta_projection}
\end{equation}
The representation $R_0$ of $G$ arises from decomposing the
symmetric product of two adjoint representations of $G$ under $G$.
Furthermore, the representation $R_0$, when branched under $SO(N)$,
is a unique representation in the above decomposition that contains
the $\boxplus$ representation of $SO(N)$.
\\
\indent From the T-tensors, it is straightforward to compute $A_1$
and $A_2$ tensors which are relevant in computing the scalar
potential via the following formulae
\begin{eqnarray}
A_1^{IJ}&=&-\frac{4}{N-2}T^{IM,JM}+\frac{2}{(N-1)(N-2)}\delta^{IJ}T^{MN,MN},\nonumber\\
A_{2j}^{IJ}&=&\frac{2}{N}T^{IJ}_{\phantom{as}j}+\frac{4}{N(N-2)}f^{M(I
m}_{\phantom{as}j}T^{J)M}_{\phantom{as}m}+\frac{2}{N(N-1)(N-2)}\delta^{IJ}f^{KL\phantom{a}m}_{\phantom{as}j}T^{KL}_{\phantom{as}m},
\nonumber \\
V&=&-\frac{4}{N}g^2\left(A_1^{IJ}A_1^{IJ}-\frac{1}{2}Ng^{ij}A_{2i}^{IJ}A_{2j}^{IJ}\right)
\end{eqnarray}
\subsection{$N=6$ gauged supergravity in three dimensions}
We now in a position to study some gaugings and their associated
vacua of $N=6$ gauged supergravity. As mentioned earlier, the scalar
manifold is the coset space $\frac{SU(4,k)}{S(U(4)\times U(k))}$
where $k$ is the number of matter multiplets.
\\
\indent To parametrize the coset representative $L$, we first
construct the global symmetry group $G=SU(4,k)$. In this paper, we
study four cases namely $k=1,2,3,4$. We use the standard $SU(4+k)$
generators in the form of generalized Gell-Mann matrices. These
matrices can be found in many text books for example \cite{stancu}.
We will denote these matrices as $c_i$, $i=1,\ldots, (4+k)^2-1$. The
non-compact form $SU(4,k)$ can be constructed from $SU(4+k)$ by the
Weyl unitarity trick. This is achieved by introducing a factor of
$i$ to each generator which is identified to be non-compact. The
maximal compact subgroup $H=S(U(4)\times U(k))\sim SU(4)\times
SU(k)\times U(1)$ can be easily identified from the standard
construction of the $SU(4+k)$ generators. All other generators are
then the non-compact ones. We have explicitly identified all
non-compact generators for the $k=4$ case in the appendix. To obtain
the non-compact generators for other values of $k$, we simply read
them off from the 32 $Y$'s of the $k=4$ case. In the $k=1$ case,
there are eight non-compact generators which are the first eight
generators, $Y_1,\ldots, Y_8$. $Y_1,\ldots, Y_{16}$ and $Y_1,\ldots,
24$ are non-compact generators of the $k=3$ and $k=4$ cases,
respectively.
\\
\indent Having identified compact and non-compact generators, we can
now use the commutation relation $\left[X^{IJ},Y^A\right]$ in
\eqref{Galgebra} to find the $f^{IJ}$ tensor whose components are
denoted by $f^{IJ}_{ij}$ or $f^{IJ}_{AB}$ in the flat basis. With
the normalization of $Y^A$ to be one, we can explicitly write
\begin{equation}
f^{IJ}_{AB}=-2\textrm{Tr}(Y^B\left[X^{IJ},Y^A\right]).
\end{equation}
It is now easy to use \textsl{Mathematica} to compute all $f^{IJ}$.
\\
\indent For the parametrization of $L$, it is very useful to use
$SU(n)$ Euler angle parametrization given in \cite{SUN_Euler}. In
\cite{SUN_Euler}, the parametrization of $SU(n+1)$ using $U(n)$
Euler angles has been given. We can apply this parametrization
directly to the coset space of the form $\frac{SU(n,1)}{SU(n)\times
U(1)}$. For the space of the form $\frac{SU(n,m)}{SU(n) \times
U(m)}$ for $m\neq 1$, we can apply the general procedure, explained
in \cite{exceptional coset}, in parametrizing a Lie group $G$ using
Euler angles of its subgroup $H$. In the present case, we
parametrize $SU(n,m)$ using $SU(n) \times SU(m)\times U(1)$ Euler
angles. We will give the explicit form of $L$ in each case in the
following sections. The detail of the parametrization can be found
in the appendix.
\\
\indent To find admissible gauge groups which can be gauged
consistently with supersymmetry, the corresponding embedding tensors
have to satisfy the two constraints \eqref{theta_quadratic} and
\eqref{theta_projection}. As mentioned above, equation
\eqref{T_projection} is equivalent to \eqref{theta_projection} for
symmetric target space. Furthermore, it can be shown that the
quadratic constraint \eqref{theta_quadratic} is equivalent to the
relation
\begin{equation}
2A_1^{IK}A_1^{KJ}-NA_2^{IKi}A_{2i}^{JK}=\frac{1}{N}\delta^{IJ}(2A_1^{KL}A_1^{KL}-NA_2^{KLi}A^{KL}_{2i}).\label{quadratic2}
\end{equation}
This equivalence has been shown explicitly for $N=16$ theory in
\cite{nicolai2}.
\\
\indent These conditions are important in searching for admissible
gauge groups. For compact gaugings, some gauge groups have been
classified in \cite{dewit}. The associated embedding tensors are
given by
\begin{equation}
\Theta=\Theta_{SO(p)}-\Theta_{SO(6-p)}+\alpha
\Theta_{SU(k)}-\frac{4\alpha(k-1)+k(p-3)}{4+k}\Theta_{U(1)}
\end{equation}
where $\alpha$, the relative coupling between $SU(k)$ and
$SO(p)\times SO(6-p)$, is a free parameter and $\Theta
_{SO(p)}-\Theta_{SO(6-p)}$ is given by
\begin{eqnarray}
\Theta_{IJ,KL}&=&\theta \delta^{KL}_{IJ}+\delta_{[I[K}\Xi_{L]J]},\\
\Xi_{IJ}&=&\left\{\begin{array}{c}
                 2\left(1-\frac{p}{N}\right)\delta_{IJ}, \qquad I\leq p \\
                 -\frac{2p}{N}\delta_{IJ},\qquad I>p
               \end{array}
\right ., \qquad \theta=\frac{2p-N}{N}\, .
\end{eqnarray}
\indent In this paper, we will study gauge groups of non-compact
type. After finding the gauge groups, we study their vacua by
studying critical points of the resulting scalar potentials. As we
have already seen, the potential can be computed from the $A_1$ and
$A_2$ tensors which can be obtained from various components of the
T-tensors. The T-tensors are computed from the embedding tensor and
$\mathcal{V}$'s from \eqref{cosetFormula}.
\\
\indent For conveniences, we repeat the stationarity condition for
finding the critical points of the scalar potential given in
\cite{dewit}
\begin{equation}
3A_1^{IK}A^{KJ}_{2j}+Ng^{kl}A^{IK}_{2k}A^{KJ}_{3lj}=0\label{extremum}
\end{equation}
where $A^{KL}_{3lj}$ is defined by
\begin{equation}
A^{IJ}_{3ij}=\frac{1}{N^2}\bigg[-2D_{(i}D_{j)}A^{IJ}_1+g_{ij}A^{IJ}_1+A^{K[I}_1f^{J]K}_{ij}+2T_{ij}\delta^{IJ}-4D_{[i}T^{IJ}_{\phantom{as}
j]}-2T_{k[i}f^{IJk}_{\phantom{asd}j]}\bigg].
\end{equation}
For supersymmetric critical points, the unbroken supersymmetries are
encoded in the condition
\begin{equation}
A_1^{IK}A_1^{KJ}\epsilon^J=
-\frac{V_0}{4g^2}\epsilon^I=\frac{1}{N}(A_1^{KJ}A_1^{KJ}-\frac{1}{2}Ng^{ij}A_{2i}^{KJ}A_{2i}^{KJ})\epsilon^I\,
.\label{a1a2condition}
\end{equation}
The amount of unbroken supersymmetries correspond to the number of
$\epsilon^I$ that are eigenvectors of $A_1^{IJ}$.
\section{Compact gauge groups and their vacua}\label{compact}
In this section, we study some vacua of $N=6$ gauged supergravity
with compact gauge groups. The gauge groups considered here are
$SO(p)\times SO(6-p)\times SU(k)\times U(1)$ for $p=0,1,2,3$. We
consider four cases, $k=1, 2, 3, 4$, separately.
\\
\indent Before going to the discussion of the scalar potentials and
their critical points, we give the R-symmetry generators $X^{IJ}$ in
terms of the generalized Gell-Mann matrices $c_i$. We emphasize here
that our convention is such that $c_i$ are anti-hermitian. The first
fifteen $c_i$ matrices are generators of $SU(4)\sim SO(6)$ for all
$k=1,2,3,4$ cases. We obtain $X^{IJ}$ from $c_1,\ldots, c_{15}$ via
the following relation
\begin{eqnarray}
X^{12}&=&\frac{1}{2}c_3+\frac{1}{2\sqrt{3}}c_8-\frac{1}{\sqrt{6}}c_{15},\qquad
X^{13}=-\frac{1}{2}(c_2+c_{14}),\qquad
X^{23}=\frac{1}{2}(c_1-c_{13}),\nonumber \\
X^{34}&=&\frac{1}{2}c_3-\frac{1}{2\sqrt{3}}c_8+\frac{1}{\sqrt{6}}c_{15},\qquad
X^{14}=\frac{1}{2}(c_1+c_{13}),\qquad X^{35}=-\frac{1}{2}(c_6+c_9),\nonumber \\
X^{56}&=&\frac{1}{\sqrt{3}}c_8+\frac{1}{\sqrt{6}}c_{15},\qquad
X^{36}=-\frac{1}{2}(c_7+c_{10}),\qquad
X^{24}=\frac{1}{2}(c_2-c_{14}),\nonumber \\
X^{45}&=&\frac{1}{2}(c_7-c_{10}),\qquad
X^{46}=\frac{1}{2}(c_9-c_6),\qquad
X^{15}=\frac{1}{2}(c_4-c_{11}),\nonumber \\
X^{16}&=&\frac{1}{2}(c_5-c_{12}),\qquad
X^{25}=\frac{1}{2}(c_5+c_{12}),\qquad
X^{26}=-\frac{1}{2}(c_4+c_{11}).
\end{eqnarray}
It is easy to verify that these generators satisfy the
$\left[X^{IJ},X^{KL}\right]$ commutator given in \eqref{Galgebra}.
\subsection{The $k=1$ case}
This is the simplest case namely there is only one matter multiplet.
The structure of the scalar manifold is very simple and contains
only eight scalars. We can study the scalar potential by
parametrizing the full scalar manifold with all eight scalars turned
on. With Euler angle parametrization given in \cite{SUN_Euler}, the
coset representative of $\frac{SU(4,1)}{SU(4)\times U(1)}$ can be
written as
\begin{equation}
L=e^{a_1c_3}e^{a_2c_2}e^{a_3c_3}e^{a_4c_5}e^{a_5c_8}e^{a_6c_{10}}e^{a_7c_{15}}e^{\frac{i}{\sqrt{2}}b_1c_{17}}\,
.\label{k1_compactL}
\end{equation}
Notice that the factor of $i$ in the $c_{17}$ matrix indicates that
the generator corresponding to $c_{17}$ is non-compact. Since $k=1$,
there is no additional factor $SU(k)$ in the gauge groups. As a
consequence, there is no free parameter $\alpha$. Together with a
very simple structure of the scalar manifold whose scalar manifold
involves only one non-compact generator, we expect the resulting
scalar potential to be very simple. We will see that this is indeed
the case.
\\
\indent With the $L$ given in \eqref{k1_compactL}, it turns out that
all gauge groups, $SO(3)\times SO(3)$ and $SO(p)\times SO(p-6)\times
U(1)$ for $p=0,1,2$, give rise to the same potential
\begin{equation}
V=-8g^2[5+3\cosh(\sqrt{2}b_1)].\label{V_k1_SO6}
\end{equation}
Notice that the potential does not depend on $a_i$'s. This is simply
because the potential is gauge invariant, and the parametrization
\eqref{k1_compactL} has some intersections with the gauge group.
\\
\indent It is clear that there is only one trivial critical point at
$b_1=0$ with $V_0=V(b_1=0)=-64g^2$. This critical point preserves
all gauge symmetries and supersymmetries. The corresponding
supersymmetries of the trivial critical point for the gauge groups
containing a factor $SO(p)\times SO(6-p)$ are given by $(p,6-p)$.
From now on, we express the number of supersymmetries, preserved by
a given critical point, with the notation $(n_-,n_+)$ where $n_-$
and $n_+$ are number of negative and positive eigenvalues of the
corresponding $A_1$ tensor that satisfy the condition
\eqref{a1a2condition}. This notation is also related to the number
of supersymmetries in the dual two dimensional CFT.
\\
\indent We now compute the scalar mass spectrum at the trivial
critical point. To do this, we go to the so-called unitary gauge in
which the eight scalars are parametrized by
\begin{equation}
L=e^{\phi_1Y_1}e^{\phi_2Y_2}e^{\phi_3Y_3}e^{\phi_4Y_4}e^{\phi_5Y_5}e^{\phi_6Y_6}e^{\phi_7Y_7}e^{\phi_8Y_8}.\label{Lk1_unitary}
\end{equation}
With this parametrization, it is easily verified that all $\phi_i$'s
are canonically normalized by computing their kinetic terms. The
full scalar potential in this gauge is given by
\begin{eqnarray}
V&=&-g^2 \left[19+3 \cosh (\sqrt{2} \phi_5) \cosh(\sqrt{2} \phi_6)+3
\cosh(\sqrt{2} \phi_7)
\cosh(\sqrt{2} \phi_8)\right .\nonumber \\
& &\left . +3 \cosh(\sqrt{2} \phi_5) \cosh(\sqrt{2} \phi_6)
\cosh(\sqrt{2} \phi_7) \cosh(\sqrt{2} \phi_8)+3 \cosh(\sqrt{2}
\phi_3)\times \
\right .\nonumber \\
& &\left .
 \cosh(\sqrt{2} \phi_4)
\left[1+\cosh(\sqrt{2} \phi_5) \cosh(\sqrt{2} \phi_6)\right]
\left[1+\cosh(\sqrt{2} \phi_7)
\cosh(\sqrt{2} \phi_8)\right]\right .\nonumber \\
& &\left .+3 \cosh(\sqrt{2} \phi_1) \cosh(\sqrt{2} \phi_2)
 \left[1+\cosh(\sqrt{2} \phi_3) \cosh(\sqrt{2} \phi_4)\right]
\times \
\right .\nonumber \\
& &\left . \left[1+\cosh(\sqrt{2} \phi_5) \cosh(\sqrt{2}
\phi_6)\right] \left[1+\cosh(\sqrt{2} \phi_7) \cosh(\sqrt{2}
\phi_8)\right]\right].\label{V_k1_Uni}
\end{eqnarray}
The mass spectrum is given in table below.
\begin{center}
\begin{tabular}{|c|c|c|}
  \hline
  $m^2L^2$ &  $(h,\bar{h})$ & $SO(6)$  \\ \hline
   $-\frac{3}{4}$& $\left(\frac{3}{4},\frac{3}{4}\right)$ & $\mathbf{4}$ \\
  \hline
  $-\frac{3}{4}$& $\left(\frac{1}{4},\frac{1}{4}\right)$ & $\bar{\mathbf{4}}$ \\
  \hline
\end{tabular}
\end{center}
In the table, we have shown the value of $m^2L^2$, the corresponding
representation under the conformal group $SO(2,2)$ labeled by
$(h,\bar{h})$ and the representation under the symmetry group
$SO(6)$. The relations between the conformal weight $\Delta$,
$(h,\bar{h})$ and $m^2L^2$ are given by
\begin{eqnarray}
\Delta &=&h+\bar{h},\nonumber \\
m^2L^2&=&\Delta(\Delta-2).
\end{eqnarray}
For scalar fields, there is also the relation $h=\bar{h}$ since the
spin $s$ is given by $h-\bar{h}$.
\\
\indent It should be noted that the potential \eqref{V_k1_SO6} seems
to give mass to only one scalar and leaves the other seven scalars
massless. However, the mass of scalars cannot be directly obtained
from this potential since in this case, the scalars are not
canonically normalized. It is like using the cartesian and polar
coordinates. If we redefine the scalars $\phi_i$'s in
\eqref{V_k1_Uni} by using one scalar $b$ and seven angle variables
$\theta_i$ for $i=1,\ldots, 7$ such that $\sum_{i=1}^7\phi_i^2=b^2$
and naively compute the mass spectrum, we will find that only $b$
gets mass. Furthermore, in terms of these redefined scalars, the
kinetic terms are no longer canonical. There is no contradiction
here since the parametrization \eqref{k1_compactL} cannot be used to
find masses of the scalars, directly.
\\
\indent It is remarkable and unexpected that even in the unitary
gauge, all compact gauge groups, in this $k=1$ case, give rise to
the same scalar potential. It is surprising, and we do not know any
deep reason for that. The scalar mass spectra for other gauge groups
are given below.
\begin{itemize}
  \item For $SO(5)\times U(1)$ gauge group, the scalar mass spectrum
at the trivial critical point is given below.
\begin{center}
\begin{tabular}{|c|c|c|}
  \hline
  $m^2L^2$ &  $(h,\bar{h})$ & $SO(5)$  \\ \hline
   $-\frac{3}{4}$& $\left(\frac{3}{4},\frac{3}{4}\right)$ & $\mathbf{4}$ \\
  \hline
  $-\frac{3}{4}$& $\left(\frac{1}{4},\frac{1}{4}\right)$ & $\mathbf{4}$ \\
  \hline
\end{tabular}
\end{center}
\item For $SO(4)\times SO(2)\times U(1)$ gauge group, the scalar mass
spectrum at the trivial critical point is given below.
\begin{center}
\begin{tabular}{|c|c|c|}
  \hline
  $m^2L^2$ &  $(h,\bar{h})$ & $SO(4)\sim SU(2)\times SU(2)$  \\ \hline
   $-\frac{3}{4}$& $\left(\frac{3}{4},\frac{3}{4}\right)$ & $(\mathbf{2},\mathbf{1})+(\mathbf{1},\mathbf{2})$ \\
  \hline
  $-\frac{3}{4}$& $\left(\frac{1}{4},\frac{1}{4}\right)$ & $(\mathbf{2},\mathbf{1})+(\mathbf{1},\mathbf{2})$ \\
  \hline
\end{tabular}
\end{center}
\item For $SO(3)\times SO(3)$ gauge group, the scalar mass spectrum
at the trivial critical point is given below.
\begin{center}
\begin{tabular}{|c|c|c|}
  \hline
  $m^2L^2$ &  $(h,\bar{h})$ & $SO(3)\times SO(3)$  \\ \hline
   $-\frac{3}{4}$& $\left(\frac{3}{4},\frac{3}{4}\right)$ & $(\mathbf{2},\mathbf{2})$ \\
  \hline
  $-\frac{3}{4}$& $\left(\frac{1}{4},\frac{1}{4}\right)$ & $(\mathbf{2},\mathbf{2})$ \\
  \hline
\end{tabular}
\end{center}
\end{itemize}

\subsection{The $k=2$ case}
For $k>1$, there is an additional independent coupling of the gauge
group $SU(k)$. The relative coupling between the gauge group $SU(k)$
and $SO(p)\times SO(6-p)$ is denoted by the parameter $\alpha$
mentioned in the previous section. Since the number of scalar fields
in this case is sixteen which is too difficult to study the scalar
potential on the full scalar manifold. We will apply the method
introduced in \cite{warner}. In this approach, we study the
potential on a submanifold of $\frac{SU(4,2)}{SU(4)\times
SU(2)\times U(1)}$ coset space. This submanifold is invariant under
a symmetry which is a subgroup of the gauge group. Using a
consequence of Schur's lemma, it has been shown in \cite{warner}
that critical points of the potential restricted on this submanifold
are critical point of the potential evaluated on the full scalar
manifold as well.
\\
\indent We find some non-trivial critical points on the submanifold
with $U(1)_{\textrm{diag}}$ symmetry. The scalars in this sector are
contained in the coset manifold $\frac{SU(2,2)}{SU(2)\times
SU(2)\times U(1)}$, so there are eight scalars to be parametrized.
For $SO(6)\times SU(2)\times U(1)$, $SO(4)\times SO(2)\times
SU(2)\times U(1)$ and $SO(3)\times SO(3)\times SU(2)\times U(1)$
gauge groups, the residual gauge symmetry $U(1)_{\textrm{diag}}$ is
generated by $X_{12}+X_{56}$ where $X_{IJ}$ are generators of
$SO(6)$ as usual. Using the Euler angle parametrization, we find
that the coset representative in this case is given by
\begin{equation}
L=e^{a_1c_{33}}e^{a_2c_{34}}e^{a_3K_3}e^{a_5M_1}e^{a_6M_2}e^{a_4M_3}e^{\frac{i}{\sqrt{2}}b_1c_{18}}
e^{\frac{i}{\sqrt{2}}b_2c_{31}}\label{k2_SO6_L}
\end{equation}
where
\begin{eqnarray}
K_3&=&\frac{1}{\sqrt{2}}\left[c_{33},c_{34}\right],\qquad
M_1=-\frac{1}{2\sqrt{2}}\left[c_{18},c_{22}\right],\nonumber \\
M_2&=&-\frac{1}{2\sqrt{2}}\left[c_{19},c_{22}\right],\qquad
M_3=\frac{1}{\sqrt{2}}\left[M_1,M_2\right].\label{data1}
\end{eqnarray}
\indent For $SO(5)\times SU(2)\times U(1)$ gauge group, the
generator $X_{56}$ is not a generator of $SO(5)$, so we cannot use
the coset representative \eqref{k2_SO6_L}. We then choose the
generator of the $U(1)_{\textrm{diag}}$ to be $X_{12}+X_{34}$. The
coset representative for this case is given by
\begin{equation}
L=e^{a_1\kappa}e^{a_2c_{14}}e^{a_3\kappa}e^{a_4c_{33}}e^{a_5c_{34}}e^{a_6\lambda}e^{\frac{i}{\sqrt{2}}b_1c_{20}}
e^{\frac{i}{\sqrt{2}}b_2c_{31}}\label{k2_SO5_L}
\end{equation}
where
\begin{equation}
\kappa=\frac{1}{\sqrt{2}}\left[c_{13},c_{14}\right], \qquad
\lambda=\frac{2}{\sqrt{10}}c_{24}-\frac{3}{\sqrt{15}}c_{35}\,
.\label{data2}
\end{equation}
We now study critical points of each gauge group, separately.
\subsubsection{$SO(6)\times SU(2)\times U(1)$ gauging}
With the coset representative \eqref{k2_SO6_L}, we find the
following potential
\begin{eqnarray}
 V&=&\frac{1}{8} g^2 \left[-222+32 (-3+2 \alpha +\alpha ^2) \cosh(\sqrt{2} b_1)-2 (1+\alpha )^2 \cosh(2 \sqrt{2} b_1)\right.\nonumber\\
 &&\left.-48 \cosh[\sqrt{2} (b_1-b_2)]-32 \alpha  \cosh[\sqrt{2} (b_1-b_2)]-16 \alpha ^2 \cosh[\sqrt{2} (b_1-b_2)]\right.\nonumber\\
 &&\left.+\cosh[2 \sqrt{2} (b_1-b_2)]+2 \alpha  \cosh[2 \sqrt{2} (b_1-b_2)]+\alpha ^2 \cosh[2 \sqrt{2} (b_1-b_2)]\right.\nonumber\\
 &&\left.-96 \cosh(\sqrt{2} b_2)+64 \alpha  \cosh(\sqrt{2} b_2)+32 \alpha ^2 \cosh(\sqrt{2} b_2)-2 \cosh(2 \sqrt{2} b_2)\right.\nonumber\\
 &&\left.-4 \alpha  \cosh(2 \sqrt{2} b_2)-2 \alpha ^2 \cosh(2 \sqrt{2} b_2)-48 \cosh[\sqrt{2} (b_1+b_2)]\right.\nonumber\\
 &&\left.-32 \alpha  \cosh[\sqrt{2} (b_1+b_2)]-16 \alpha ^2 \cosh[\sqrt{2} (b_1+b_2)]+\cosh[2 \sqrt{2} (b_1+b_2)]\right.\nonumber\\
 &&\left.+2 \alpha  \cosh[2 \sqrt{2} (b_1+b_2)]+\alpha ^2 \cosh[2 \sqrt{2} (b_1+b_2)]-60 \alpha -30 \alpha ^2\right].\label{potential_k2_SO6}
\end{eqnarray}
We find three critical points shown in Table \ref{table1}.
\TABLE{\begin{tabular}{|c|c|c|c|c|}
  \hline
  &$b$ &  $V_0$ & unbroken SUSY & unbroken gauge symmetry \\ \hline
  I &$0$ & $-64g^2$ & $(6,0)$ & $SO(6)\times SU(2)\times U(1)$ \\
  II &$\frac{1}{\sqrt{2}}\cosh^{-1}\left(\frac{\alpha-1}{\alpha+1}\right)$, $\alpha<-1$
  &  $-\frac{16g^2(1+2\alpha)^2}{(1+\alpha)^2}$ & $(4,0)$ & $SU(2)\times SU(2)\times U(1)_{\textrm{diag}}$ \\
  III &$\frac{1}{\sqrt{2}}\cosh^{-1}\left(\frac{\alpha+3}{\alpha+1}\right)$, $\alpha>-1$ &
  $-\frac{16g^2(3+2\alpha)^2}{(1+\alpha)^2}$ & $(2,0)$ & $SU(2)\times SU(2)\times U(1)_{\textrm{diag}}$ \\
  \hline
\end{tabular}\caption{Critical points of $SO(6)\times SU(2)\times U(1)$ gauging for the $k=2$ case.}}\label{table1}
$V_0$ is the value of the potential at the corresponding critical
point or the cosmological constant. At all critical points,
$b_1=b_2=b$ where $b$ is shown in the table. The scalar mass
spectrum at the trivial critical point is given below.
\begin{center}
\begin{tabular}{|c|c|c|}
  \hline
  $m^2L^2$ &  $(h,\bar{h})$ & $SO(6)\times SU(2)$  \\ \hline
   $-\frac{3}{4}$& $\left(\frac{3}{4},\frac{3}{4}\right)$ & $(\mathbf{4},\mathbf{2})$ \\
  \hline
  $-\frac{3}{4}$& $\left(\frac{1}{4},\frac{1}{4}\right)$ & $(\bar{\mathbf{4}},\mathbf{2})$ \\
  \hline
\end{tabular}
\end{center}

\subsubsection{$SO(5)\times SU(2)\times U(1)$ gauging}
In this gauge group, the coset representative is given in
\eqref{k2_SO5_L}. The potential turns out to be the same as in
equation \eqref{potential_k2_SO6}. The critical points are of course
similar to those given in Table \ref{table1}. The amounts of
supersymmetries for critical points I, II and III are $(5,1)$,
$(4,1)$ and $(1,0)$, respectively. The positions and values of the
cosmological constant are the same as those in the $SO(6)\times
SU(2)\times U(1)$ gauging. However, unbroken gauge symmetry of II
and III critical points is given by $SU(2)_{\textrm{diag}}\times
SU(2)$. The $SU(2)_{\textrm{diag}}$ is a diagonal subgroup of the
$SU(2)$ and one of the $SU(2)$ in the $SO(4)\subset SO(5)$. The
scalar mass spectrum at the trivial critical point is given below.
\begin{center}
\begin{tabular}{|c|c|c|}
  \hline
  $m^2L^2$ &  $(h,\bar{h})$ & $SO(5)\times SU(2)$  \\ \hline
   $-\frac{3}{4}$& $\left(\frac{3}{4},\frac{3}{4}\right)$ & $(\mathbf{4},\mathbf{2})$ \\
  \hline
  $-\frac{3}{4}$& $\left(\frac{1}{4},\frac{1}{4}\right)$ & $(\mathbf{4},\mathbf{2})$ \\
  \hline
\end{tabular}
\end{center}

\subsubsection{$SO(4)\times SO(2)\times SU(2)\times U(1)$ gauging}
The coset representative for this case is given in \eqref{k2_SO6_L}.
We find the potential
\begin{eqnarray}
 V&=&-\frac{1}{8} g^2 \left[192+30 \alpha ^2-32 (-4+\alpha ^2) \cosh(\sqrt{2} b_1)+2 \alpha ^2 \cosh(2 \sqrt{2} b_1)\right.\nonumber\\
 &&\left.+32 \cosh[\sqrt{2} (b_1-b_2)]+16 \alpha ^2 \cosh[\sqrt{2} (b_1-b_2)]-\alpha ^2 \cosh[2 \sqrt{2} (b_1-b_2)]\right.\nonumber\\
 &&\left.+128 \cosh(\sqrt{2} b_2)-32 \alpha ^2 \cosh(\sqrt{2} b_2)+2 \alpha ^2 \cosh(2 \sqrt{2} b_2)\right.\nonumber\\
 &&\left.+32 \cosh[\sqrt{2} (b_1+b_2)]+16 \alpha ^2 \cosh[\sqrt{2} (b_1+b_2)]
\right.\nonumber\\
 &&\left.
 -\alpha ^2 \cosh[2 \sqrt{2} (b_1+b_2)]\right].\label{potential_k2_SO4}
\end{eqnarray}
As in the $SO(6)\times SU(2)\times U(1)$ gauging, we find three
critical points shown in Table \ref{table2}.
\TABLE{\begin{tabular}{|c|c|c|c|c|}
  \hline
  &$b$ &  $V_0$ & unbroken SUSY & unbroken gauge symmetry \\ \hline
  I$'$ &$0$ & $-64g^2$ & $(4,2)$ & $SO(4)\times SO(2)\times SU(2)\times U(1)$ \\
  II$'$ &$\frac{1}{\sqrt{2}}\cosh^{-1}\left(\frac{\alpha+2}{\alpha}\right)$, $\alpha>0$ &
  $-\frac{16g^2(1+2\alpha)^2}{\alpha^2}$ & $(2,2)$ & $U(1)_{\textrm{diag}}\times U(1)\times U(1)$ \\
  III$'$ &$\frac{1}{\sqrt{2}}\cosh^{-1}\left(\frac{\alpha-2}{\alpha}\right)$, $\alpha<0$ &
  $-\frac{16g^2(1-2\alpha)^2}{\alpha^2}$ & $(2,0)$ & $U(1)_{\textrm{diag}}\times U(1)\times U(1)$ \\
  \hline
\end{tabular}\caption{Critical points of $SO(4)\times SO(2)\times SU(2)\times U(1)$ gauging for the $k=2$ case.}}\label{table2}
As in Table \ref{table1}, $b_1=b_2=b$ at the critical points. The
scalar mass spectrum at the trivial critical point is given below.
\begin{center}
\begin{tabular}{|c|c|c|}
  \hline
  $m^2L^2$ &  $(h,\bar{h})$ & $SO(4)\times SU(2)\sim SU(2)\times SU(2)\times SU(2)$  \\ \hline
   $-\frac{3}{4}$& $\left(\frac{3}{4},\frac{3}{4}\right)$ & $(\mathbf{1},\mathbf{2},\mathbf{2})+(\mathbf{2},\mathbf{1},\mathbf{2})$ \\
  \hline
  $-\frac{3}{4}$& $\left(\frac{1}{4},\frac{1}{4}\right)$ & $(\mathbf{1},\mathbf{2},\mathbf{2})+(\mathbf{2},\mathbf{1},\mathbf{2})$ \\
  \hline
\end{tabular}
\end{center}

\subsubsection{$SO(3)\times SO(3)\times SU(2)\times U(1)$ gauging}
With the same coset representative as in the $SO(4)\times
SO(2)\times SU(2)\times U(1)$ gauging, this gauge group gives rise
to the same potential as in \eqref{potential_k2_SO4}. The three
critical points are characterized by the same parameters as those in
Table \ref{table2} except for the critical points I$'$, II$'$ and
III$'$ having $(3,3)$, $(1,2)$ and $(2,1)$ supersymmetries,
respectively, and the unbroken gauge symmetry for critical points
II$'$ and III$'$ being $U(1)_{\textrm{diag}}\times U(1)$. The scalar
mass spectrum at the trivial critical point is given below.
\begin{center}
\begin{tabular}{|c|c|c|}
  \hline
  $m^2L^2$ &  $(h,\bar{h})$ & $SO(3)\times SO(3)\times SU(2)$  \\ \hline
   $-\frac{3}{4}$& $\left(\frac{3}{4},\frac{3}{4}\right)$ & $(\mathbf{2},\mathbf{2},\mathbf{2})$ \\
  \hline
  $-\frac{3}{4}$& $\left(\frac{1}{4},\frac{1}{4}\right)$ & $(\mathbf{2},\mathbf{2},\mathbf{2})$ \\
  \hline
\end{tabular}
\end{center}
\indent Notice that for $SO(6)\times SU(2)\times U(1)$ and
$SO(5)\times SU(2)\times U(1)$ gaugings, there is still one
non-trivial critical point when $\alpha=0$, in the limit where
$SU(k)$ decouples. This is not the case for $SO(3)\times SO(3)\times
SU(2)\times U(1)$ and $SO(4)\times SO(2)\times SU(2)\times U(1)$
gaugings. These two gauge groups do not have any non-trivial
critical points in our parametrization when $\alpha=0$. As we will
see, the same pattern will appear again in the $k=3,4$ cases. This
could be an artifact of our specific parametrization, but after
various attempts, this seems to be the only one that gives rise to
non-trivial critical points while the complication of the
computation is still controllable. Furthermore, we refrain from
giving the scalar mass spectrum at the non-trivial critical points
because the corresponding calculation is more complicated.
\subsection{The $k=3$ case}
There are 24 scalars in this case. We choose the scalar submanifold
with the same residual symmetry as in the previous case,
$U(1)_{\textrm{diag}}$. There are again two coset representatives,
one for $SO(5)\times SU(3)\times U(1)$ gauging and the other one for
$SO(p)\times SO(6-p)\times SU(3)\times U(1)$ for $p=3,4,6$. Each
coset representative contains 12 scalars parametrizing
$\frac{SU(2,3)}{SU(3)\times SU(2)\times U(1)}$ coset.
\\
\indent The coset representative invariant under $X_{12}+ X_{56}$ is
given by
\begin{equation}
L=e^{a_1\Lambda_3}e^{a_2c_{34}}e^{a_3\Lambda_3}e^{a_4c_{45}}e^{a_5\Lambda_8}
e^{a_6\Lambda_3}e^{a_7c_{34}}e^{a_8\Lambda_3}e^{a_9M_3}e^{a_{10}M_2}
e^{\frac{i}{\sqrt{2}}b_1c_{18}} e^{\frac{i}{\sqrt{2}}b_1c_{42}}
\end{equation}
where
\begin{equation}
\Lambda_3=\frac{2}{\sqrt{10}}c_{24}-\frac{3}{\sqrt{15}}c_{35},\qquad
\Lambda_8=\frac{2}{\sqrt{30}}c_{24}+\frac{2}{3\sqrt{5}}c_{35}-\frac{7}{3\sqrt{7}}c_{48},\label{Lambda}
\end{equation}
and $M_2$ and $M_3$ are given in \eqref{data1}. This $L$ will be
used for the $SO(p)\times SO(6-p)\times SU(3)\times U(1)$ for $p=3,
4, 6$ gaugings.
\\
\indent The coset representative for the $SO(5)\times SU(3)\times
U(1)$ gauging is invariant under $X_{12}+ X_{34}$ and given by
\begin{equation}
L=e^{a_1\Lambda_3}e^{a_2c_{34}}e^{a_3\Lambda_3}e^{a_4c_{45}}e^{a_5\Lambda_8}
e^{a_6\Lambda_3}e^{a_7c_{34}}e^{a_8\Lambda_3}e^{a_9k}e^{a_{10}c_{14}}
e^{\frac{i}{\sqrt{2}}b_1c_{20}} e^{\frac{i}{\sqrt{2}}b_1c_{31}}
\end{equation}
where $k$ is given in \eqref{data2} and $\Lambda_3$ and $\Lambda_8$
are given in \eqref{Lambda}.
\\
\indent The scalar potential and structure of the critical points
turn out to be similar to what we have found in the previous case.
The only difference lies in the residual gauge symmetry of the
critical points. The trivial critical point at $L=\mathbf{I}$ always
preserves the full gauge symmetry. The non-trivial points II, III,
II$'$ and III$'$ have the following residual symmetries.
\begin{itemize}
  \item  $SO(6)\times SU(3)\times U(1)$ gauge group: $SU(2)\times SU(2)\times U(1)\times U(1)$
  \item $SO(5)\times SU(3)\times U(1)$ gauge group: $SU(2)\times SU(2)\times U(1)$
  \item $SO(4)\times SO(3)\times SU(2)\times U(1)$ gauge group: $U(1)\times U(1)\times U(1)\times U(1)$
  \item $SO(3)\times SO(3)\times SU(3)\times U(1)$ gauge group: $U(1)\times U(1)\times U(1)$
\end{itemize}
Other properties are the same as those given in Table \ref{table1}
and \ref{table2}. The scalar mass spectra are given below.
\begin{itemize}
\item For $SO(6)\times SU(3)\times U(1)$ gauge group, the scalar mass
spectrum at the trivial critical point is given below.
\begin{center}
\begin{tabular}{|c|c|c|}
  \hline
  $m^2L^2$ &  $(h,\bar{h})$ & $SO(6)\times SU(3)$  \\ \hline
   $-\frac{3}{4}$& $\left(\frac{3}{4},\frac{3}{4}\right)$ & $(\mathbf{4},\bar{\mathbf{3}})$ \\
  \hline
  $-\frac{3}{4}$& $\left(\frac{1}{4},\frac{1}{4}\right)$ & $(\bar{\mathbf{4}},\mathbf{3})$ \\
  \hline
\end{tabular}
\end{center}
\item For $SO(5)\times SU(3)\times U(1)$ gauge group, the scalar mass
spectrum at the trivial critical point is given below.
\begin{center}
\begin{tabular}{|c|c|c|}
  \hline
  $m^2L^2$ &  $(h,\bar{h})$ & $SO(5)\times SU(3)$  \\ \hline
   $-\frac{3}{4}$& $\left(\frac{3}{4},\frac{3}{4}\right)$ & $(\mathbf{4},\bar{\mathbf{3}})$ \\
  \hline
  $-\frac{3}{4}$& $\left(\frac{1}{4},\frac{1}{4}\right)$ & $(\mathbf{4},\mathbf{3})$ \\
  \hline
\end{tabular}
\end{center}
\item For $SO(4)\times SO(2)\times SU(3)\times U(1)$ gauge group, the
scalar mass spectrum at the trivial critical point is given below.
\begin{center}
\begin{tabular}{|c|c|c|}
  \hline
  $m^2L^2$ &  $(h,\bar{h})$ & $SO(4)\times SU(3)\sim SU(2)\times SU(2)\times SU(3)$  \\ \hline
   $-\frac{3}{4}$& $\left(\frac{3}{4},\frac{3}{4}\right)$ & $(\mathbf{2},\mathbf{1},\bar{\mathbf{3}})+(\mathbf{1},\mathbf{2},\bar{\mathbf{3}})$ \\
  \hline
  $-\frac{3}{4}$& $\left(\frac{1}{4},\frac{1}{4}\right)$ & $(\mathbf{1},\mathbf{2},\mathbf{3})+(\mathbf{2},\mathbf{1},\mathbf{3})$ \\
  \hline
\end{tabular}
\end{center}
\item For $SO(3)\times SO(3)\times SU(3)\times U(1)$ gauge group, the
scalar mass spectrum at the trivial critical point is given below.
\begin{center}
\begin{tabular}{|c|c|c|}
  \hline
  $m^2L^2$ &  $(h,\bar{h})$ & $SO(3)\times SO(3)\times SU(3)$  \\ \hline
   $-\frac{3}{4}$& $\left(\frac{3}{4},\frac{3}{4}\right)$ & $(\mathbf{2},\mathbf{2},\bar{\mathbf{3}})$ \\
  \hline
  $-\frac{3}{4}$& $\left(\frac{1}{4},\frac{1}{4}\right)$ & $(\mathbf{2},\mathbf{2},\mathbf{3})$ \\
  \hline
\end{tabular}
\end{center}
\end{itemize}

\subsection{The $k=4$ case}
In this case, there are 32 scalars which are too difficult to carry
out the calculation in the $U(1)_{\textrm{diag}}$ invariant sector.
We overcome this issue by requiring more residual symmetry namely
introducing an $SU(2)\subset SU(4)$ factor. The scalar manifold we
need to parametrize is now invariant under
$U(1)_{\textrm{diag}}\times SU(2)$. Notice that in this
parametrization, we do not have the limit when $\alpha\rightarrow 0$
anymore since the residual gauge symmetry of the scalar submanifold
contains a subgroup of the $SU(k)=SU(4)$. With this extra factor of
$SU(2)$, we are left with 8 scalars. The coset representatives for
the $SO(p)\times SO(6-p)\times SU(4)\times U(1)$ for $p=3, 4, 6$,
and $SO(5)\times SU(4)\times U(1)$ gaugings are the same as those
used in the $k=2$ case and given in \eqref{k2_SO6_L} and
\eqref{k2_SO5_L}, respectively.
\\
\indent The scalar potential and structure of the critical points
are the same as the $k=2$ case except for unbroken gauge symmetries
of the non-trivial critical points. The non-trivial points II, III,
II$'$ and III$'$ have the following residual symmetries.
\begin{itemize}
  \item  $SO(6)\times SU(4)\times U(1)$ gauge group: $SU(2)\times SU(2)\times SU(2) \times U(1)\times U(1)$
  \item $SO(5)\times SU(4)\times U(1)$ gauge group: $SU(2)\times SU(2)\times SU(2)$
  \item $SO(4)\times SO(2)\times SU(4)\times U(1)$ gauge group: $SU(2)\times U(1)\times U(1)\times U(1)\times U(1)$
  \item $SO(3)\times SO(3)\times SU(4)\times U(1)$ gauge group: $SU(2)\times U(1)\times U(1)\times U(1)$
\end{itemize}
Other properties are the same as those given in Table \ref{table1}
and \ref{table2}. The scalar mass spectra are given below.
\begin{itemize}
\item For $SO(6)\times SU(4)\times U(1)$ gauge group, the scalar mass
spectrum at the trivial critical point is given below.
\begin{center}
\begin{tabular}{|c|c|c|}
  \hline
  $m^2L^2$ &  $(h,\bar{h})$ & $SO(6)\times SU(4)$  \\ \hline
   $-\frac{3}{4}$& $\left(\frac{3}{4},\frac{3}{4}\right)$ & $(\mathbf{4},\bar{\mathbf{4}})$ \\
  \hline
  $-\frac{3}{4}$& $\left(\frac{1}{4},\frac{1}{4}\right)$ & $(\bar{\mathbf{4}},\mathbf{4})$ \\
  \hline
\end{tabular}
\end{center}
\item For $SO(5)\times SU(4)\times U(1)$ gauge group, the scalar mass
spectrum at the trivial critical point is given below.
\begin{center}
\begin{tabular}{|c|c|c|}
  \hline
  $m^2L^2$ &  $(h,\bar{h})$ & $SO(5)\times SU(4)$  \\ \hline
   $-\frac{3}{4}$& $\left(\frac{3}{4},\frac{3}{4}\right)$ & $(\mathbf{4},\bar{\mathbf{4}})$ \\
  \hline
  $-\frac{3}{4}$& $\left(\frac{1}{4},\frac{1}{4}\right)$ & $(\mathbf{4},\mathbf{4})$ \\
  \hline
\end{tabular}
\end{center}
\item For $SO(4)\times SO(2)\times SU(4)\times U(1)$ gauge group, the
scalar mass spectrum at the trivial critical point is given below.
\begin{center}
\begin{tabular}{|c|c|c|}
  \hline
  $m^2L^2$ &  $(h,\bar{h})$ & $SO(4)\times SU(4)\sim SU(2)\times SU(2)\times SU(4)$  \\ \hline
   $-\frac{3}{4}$& $\left(\frac{3}{4},\frac{3}{4}\right)$ & $(\mathbf{2},\mathbf{1},\bar{\mathbf{4}})+(\mathbf{1},\mathbf{2},\bar{\mathbf{4}})$ \\
  \hline
  $-\frac{3}{4}$& $\left(\frac{1}{4},\frac{1}{4}\right)$ & $(\mathbf{1},\mathbf{2},\mathbf{4})+(\mathbf{2},\mathbf{1},\mathbf{4})$ \\
  \hline
\end{tabular}
\end{center}
\item For $SO(3)\times SO(3)\times SU(4)\times U(1)$ gauge group, the
scalar mass spectrum at the trivial critical point is given below.
\begin{center}
\begin{tabular}{|c|c|c|}
  \hline
  $m^2L^2$ &  $(h,\bar{h})$ & $SO(3)\times SO(3)\times SU(4)$  \\ \hline
   $-\frac{3}{4}$& $\left(\frac{3}{4},\frac{3}{4}\right)$ & $(\mathbf{2},\mathbf{2},\bar{\mathbf{4}})$ \\
  \hline
  $-\frac{3}{4}$& $\left(\frac{1}{4},\frac{1}{4}\right)$ & $(\mathbf{2},\mathbf{2},\mathbf{4})$ \\
  \hline
\end{tabular}
\end{center}
\end{itemize}

\section{Non-compact gauge groups and their
vacua}\label{non_compact} In this section, we find some non-compact
gauge groups by using the consistency conditions given in section
\ref{3DSUGRA}. For a given gauge group to be admissible, its
embedding tensor has to satisfy the conditions
\eqref{theta_projection} and \eqref{theta_quadratic}, or
equivalently, the T-tensors satisfy the conditions
\eqref{T_projection} and \eqref{quadratic2}. As in the compact case,
we study the four cases corresponding to $k=1,2,3,4$, separately. In
the results given below, we first give the embedding tensor of each
gauge group followed by the study of the corresponding critical
points. Since the scalars corresponding to non-compact generators of
the gauge group will drop out from the scalar potential, the number
of scalars to be parametrized given below always means the number of
scalars described by non-compact generators outside the gauge group.
At the trivial critical point where all scalars are zero, the gauge
group is broken down to its maximal compact subgroup which
constitutes the residual symmetry of the associated critical point.
Furthermore, this point preserves all supersymmetries namely $N=6$
in three dimensions. As can be seen from \eqref{a1a2condition},
supersymmetric critical points are possible only for AdS and
Minkowski critical points. It is convenient to express the number of
supersymmetries in the case of AdS critical points in the two
dimensional language of the corresponding dual CFT's in the form
$(n_-,n_+)$ as in the compact gaugings.
\subsection{The $k=1$ case}
In this case, the global symmetry group is $SU(4,1)$. We find that
the following subgroups can be gauged:
\begin{eqnarray}
SU(3,1)\times U(1)&:&\qquad \Theta= \Theta_{SU(3,1)} - \frac{3}{5}
\Theta_{U(1)} \\
SU(2,1)\times SU(2) \times U(1)&:& \qquad \Theta= \Theta_{SU(2,1)}-
\Theta_{SU(2)} - \frac{1}{5} \Theta_{U(1)}\\
SU(1,1)\times SU(3) \times U(1)&:&\qquad  \Theta=\Theta_{SU(1,1)}-
\Theta_{SU(3)} + \frac{1}{5} \Theta_{U(1)}\, .
\end{eqnarray}
Note that for gauge groups which are different real forms of the
same complex group, the relative couplings between each factor of
the full gauge group in the embedding tensor are the same. So,
$SU(2,1)\times SU(2) \times U(1)$ and $SU(1,1)\times SU(3) \times
U(1)$ have the same embedding tensor up to an irrelevant overall
constant which, in this case, is just a minus sign. This constant
depends on the normalization of the gauge generators. This fact has
been mentioned before in \cite{nicolai2} for admissible non-compact
gauge groups of the $N=16$ theory. We now study scalar potentials
for these three gauge groups.
\subsubsection{$SU(3,1)\times U(1)$ gauging}
There are six non-compact generators in the gauge group. Scalars
corresponding to these generators will drop out from the potential,
so we only need to parametrize $L$ with the remaining two scalars.
These correspond to non-compact generators of $SU(1,1)\subset
SU(3,1)$. We then choose the parametrization
\begin{equation}
L=e^{aX}e^{\frac{i}{\sqrt{2}}bc_{16}}e^{-aX},\qquad
X=-\frac{1}{\sqrt{2}}\left[c_{16},c_{17}\right].
\end{equation}
The potential is given by
\begin{equation}
V=8g^2(3\cosh(\sqrt{2}b)-5).
\end{equation}
There is no non-trivial critical point. At $b=0$, we find
$V_0=-16g^2$, and the $N=(0,6)$ supersymmetry is preserved at this
critical point. The scalar mass spectrum is given in the table
below.
\begin{center}
\begin{tabular}{|c|c|c|}
  \hline
  $m^2L^2$ &  $(h,\bar{h})$ & $SU(3)$  \\ \hline
   $3$& $\left(\frac{3}{2},\frac{3}{2}\right)$ & $2\times \mathbf{1}$ \\
  \hline
\end{tabular}
\end{center}
The six massless scalars which drop out from the potential as
mentioned before correspond to the Goldstone bosons. As a result, at
this critical point, there will be six massive vector fields
corresponding to the broken generators of $SU(3,1)$. This fact has
also been emphasized in \cite{N16Vacua}. For the supersymmetric
right moving sector, the R-symmetry in this case is a $U(3)\sim
SU(3)\times U(1)\subset SU(4)$ under which the supercharges
transform as $\mathbf{3}+\bar{\mathbf{3}}$. The corresponding
superconformal algebra is given by $SU(1,1|3)$ \cite{SC}. The
$SU(1,1)\sim SO(2,1) \sim SL(2,\mathbb{R})$ is a part of $AdS_3$
isometry group $SO(2,2)\sim SO(2,1)\times SO(2,1)$. The twelve
supercharges transform under the full bosonic subalgebra
$SU(1,1)\times SU(3)\times U(1)$ as
$(\mathbf{2},\mathbf{3})+(\mathbf{2},\bar{\mathbf{3}})$. We can also
write the full background isometry of the $(0,6)$ supersymmetric
$AdS_3$ vacuum as $SU(1,1)\times SU(1,1|3)$ in which the $SU(1,1)$
corresponds to the left moving part without any supersymmetry. On
the other hand, if the full $SU(4)\sim SO(6)$ R-symmetry is
preserved, the associated superconformal algebra will be
$OSp(6|2,\mathbb{R})$.

\subsubsection{$SU(2,1)\times SU(2) \times U(1)$ gauging}
There are four relevant scalars described by non-compact generators
of $SU(2,1)\subset SU(4,1)$. We emphasize here that this $SU(2,1)$
subgroup is not the same as that appears in the gauge group. The
parametrization of $L$ is
\begin{equation}
L=
e^{a_1q_1}e^{a_2q_2}e^{a_3q_3}e^{-\frac{i}{\sqrt{2}}bc_{16}}e^{-a_3q_3}e^{-a_2q_2}e^{-a_1q_1}
\end{equation}
where
\begin{equation}
q_i=\frac{1}{2}c_i\, .\label{qi_def}
\end{equation}
We find the potential
\begin{equation}
V=8 g^2 \left[-1+\cosh(\sqrt{2} b)\right]
\end{equation}
which does not admit any non-trivial critical points. The trivial
critical point is given by $b=0$ with $V_0=0$ and preserves $N=6$
supersymmetry. The scalar mass spectrum is given in the table below.
\begin{center}
\begin{tabular}{|c|c|c|}
  \hline
  $m^2$ &  $SU(2)\times SU(2)$  \\ \hline
   $16g^2$&  $2\times (\mathbf{1},\mathbf{2})$ \\
  \hline
\end{tabular}
\end{center}

\subsubsection{$SU(1,1)\times SU(3) \times U(1)$ gauging}
There are six scalars in $L$. They correspond to non-compact
generators of $SU(3,1)\subset SU(4,1)$. With $U(3)$ Euler angles, we
can parametrize $L$ as
\begin{equation}
L=
e^{a_1c_3}e^{a_2c_2}e^{a_3c_3}e^{a_4c_5}e^{a_5c_8}e^{\frac{i}{\sqrt{2}}bc_{17}}\,
.
\end{equation}
We find the scalar potential
\begin{equation}
V=-8 g^2 \left(1+\cosh(\sqrt{2} b)\right).
\end{equation}
At $b=0$, there is a trivial critical point with $V_0=-16g^2$. The
scalar mass spectrum is given in the table below.
\begin{center}
\begin{tabular}{|c|c|c|}
  \hline
  $m^2L^2$ &  $(h,\bar{h})$ & $SU(3)$  \\ \hline
   $-1$& $\left(\frac{1}{2},\frac{1}{2}\right)$ & $\mathbf{3}+\bar{\mathbf{3}}$ \\
  \hline
\end{tabular}
\end{center}
The trivial point is invariant under $(0,6)$ supersymmetry and
$SU(3)\times U(1)\times U(1)$ with the first two factors being a
subgroup of the $SU(4)$ R-symmetry. The corresponding superconformal
group is again $SU(1,1|3)$.

\subsection{The $k=2$ case}
In this case, the global symmetry group is $SU(4,2)$. We find that
the following gauge groups are admissible:
\begin{eqnarray}
SU(3,2)\times U(1)&:& \qquad \Theta= \Theta_{SU(3,2)} - \frac{2}{3}
\Theta_{U(1)}\\
SU(2,2)\times SU(2) \times U(1)&:& \qquad \Theta= \Theta_{SU(2,2)}-
\Theta_{SU(2)} - \frac{1}{3} \Theta_{U(1)}\\
SU(1,2)\times SU(3)&:& \qquad \Theta=\Theta_{SU(1,2)}-\Theta_{SU(3)}\\
SU(3,1)\times SU(1,1) \times U(1)&:&\qquad \Theta=\Theta_{SU(3,1)}
-\Theta_{SU(1,1)}- \frac{1}{3} \Theta_{U(1)}\\
SU(2,1)\times SU(2,1)&:& \qquad
\Theta=\Theta_{SU(2,1)}-\Theta_{SU(2,1)}\\
SU(4,1)\times U(1)&:&\qquad  \Theta=\Theta_{SU(4,1)} - \frac{2}{3}
\Theta_{U(1)}\, .
\end{eqnarray}
We now compute the associated scalar potentials.
\subsubsection{$SU(3,2)\times U(1)$ gauging}
There are four relevant scalars parametrizing the manifold
$\frac{SU(2,1)}{SU(2)\times U(1)}$ whose coset representative is
given by
\begin{equation}
L= e^{a_1Q_1}e^{a_2Q_2}e^{a_3Q_3}e^{\frac{i}{\sqrt{2}}bc_{16}}
\end{equation}
where
\begin{equation}
Q_1= \frac{1}{2}c_{33},\qquad Q_2 = \frac{1}{2}c_{34},\qquad Q_3=
\frac{1}{2}\left(\frac{2}{\sqrt{10}}c_{24}-\frac{3}{\sqrt{15}}c_{35}\right).\label{Qi_def}
\end{equation}
We obtain the potential
\begin{equation}
V=8 g^2 \left(-5+3 \cosh(\sqrt{2} b)\right)
\end{equation}
The only one critical point is given by $b=0$ with $(0,6)$
supersymmetry and $V_0=-16g^2$. The corresponding superconformal
algebra is given by $SU(1,1|3)$ with the R-symmetry part
$SU(3)\times U(1)$ being a subgroup of the $SU(4)$ R-symmetry. The
scalar mass spectrum is given in the table below.
\begin{center}
\begin{tabular}{|c|c|c|}
  \hline
  $m^2L^2$ &  $(h,\bar{h})$ & $SU(3)\times SU(2)$  \\ \hline
   $3$& $\left(\frac{3}{2},\frac{3}{2}\right)$ & $2\times (\mathbf{1},\mathbf{2})$ \\
  \hline
\end{tabular}
\end{center}

\subsubsection{$SU(2,2)\times SU(2) \times U(1)$ gauging}
There are eight scalars which do not correspond to non-compact
generators of the gauge group. These scalars form the smaller coset
space $\frac{SU(2,2)}{SU(2)\times SU(2)\times U(1)}$. The coset
representative takes the form
\begin{equation}
L= e^{a_1P_1}e^{a_2P_2} e^{a_3P_3}e^{a_4Q_1} e^{a_5Q_2}e^{a_6Q_3}
e^{\frac{i}{\sqrt{2}}b_1c_{16}}e^{\frac{i}{\sqrt{2}}b_2c_{27}},
\end{equation}
where $P_i=\frac{1}{2}c_i$ and $Q_i$'s are given in \eqref{Qi_def}.
We find the potential
\begin{equation}
V=-8 g^2 \left[3+\cosh(\sqrt{2} b_1) (-2+\cosh(\sqrt{2} b_2))-2
\cosh(\sqrt{2} b_2)\right]
\end{equation}
There are two critical points:
\begin{itemize}
\item at $b_1=b_2=0$ with $V_0=0$ and $N=6$ supersymmetry
\item at $b_1=b_2=\frac{1}{\sqrt{2}}\cosh^{-1}{2}$ with $V_0=8g^2$. This
critical point is invariant under $SU(2)_{\textrm{diag}}\times
SU(2)\times U(1)$ symmetry. The $SU(2)_{\textrm{diag}}$ is a
diagonal subgroup of the $SU(2)$ factor in the gauge group and one
of the $SU(2)$'s in $SU(2,2)$.
\end{itemize}
The scalar mass spectrum at the trivial critical point is given in
the table below.
\begin{center}
\begin{tabular}{|c|c|}
  \hline
  $m^2$ &   $SU(2)\times SU(2)\times SU(2)$  \\ \hline
   $16g^2$&  $2\times (\mathbf{1},\mathbf{2},\mathbf{2})$ \\
  \hline
\end{tabular}
\end{center}

\subsubsection{$SU(1,2)\times SU(3)$ gauging}
There are twelve relevant scalars in the coset representative which
takes the form
\begin{equation}
L= e^{a_1c_3}e^{a_2c_2} e^{a_3 c_3}e^{a_4 c_5}
e^{\frac{1}{\sqrt{3}}a_5c_8}e^{a_6c_3} e^{a_7c_2}e^{a_8c_3}
e^{a_9Q_3} e^{a_{10} Q_2 }
e^{\frac{i}{\sqrt{2}}b_1c_{16}}e^{\frac{i}{\sqrt{2}}b_2c_{27}},
\end{equation}
where $Q_i$'s are given in \eqref{Qi_def}. The scalars correspond to
non-compact generators of $SU(3,2)\subset SU(4,2)$. The potential is
found to be
\begin{equation}
V=-8 g^2 \left[1+\cosh(\sqrt{2} b_1) \cosh(\sqrt{2} b_2)\right].
\end{equation}
There is a trivial critical point at $b_1=b_2=0$ with $V_0=-16g^2$,
$(0,6)$ supersymmetry and superconformal algebra $SU(1,1|3)$. The
scalar mass spectrum is given in the table below.
\begin{center}
\begin{tabular}{|c|c|c|}
  \hline
  $m^2L^2$ &  $(h,\bar{h})$ & $SU(3)\times SU(2)$  \\ \hline
   $-1$& $\left(\frac{1}{2},\frac{1}{2}\right)$ & $(\mathbf{3},\mathbf{2})+(\bar{\mathbf{3}},\mathbf{2})$ \\
  \hline
\end{tabular}
\end{center}

\subsubsection{$SU(3,1)\times SU(1,1) \times U(1)$ gauging}
There are eight relevant scalars encoded in the
$\frac{SU(1,1)}{U(1)}\times \frac{SU(3,1)}{SU(3)\times U(1)}$ coset.
The coset representative is parametrized by
\begin{equation}
L=
e^{-\frac{1}{2\sqrt{2}}a_1[c_{16},c_{17}]}e^{\frac{i}{\sqrt{2}}b_1c_{16}}e^{a_2w_3}e^{a_3w_2}
e^{a_4w_3}e^{a_5w_5}e^{\frac{1}{\sqrt{3}}a_6w_8}e^{\frac{i}{\sqrt{2}}b_2c_{28}},
\end{equation}
where
\begin{eqnarray}
w_2&=&\frac{1}{2}c_{7},\qquad
w_3=-\frac{1}{4}(c_3-\sqrt{3}c_8),\nonumber
\\
w_5&=&\frac{1}{2}c_{12},\qquad
w_8=\frac{1}{4}(\sqrt{3}c_3+c_8-4\sqrt{2}c_{15}).\label{wi_def}
\end{eqnarray}
The potential is given by
\begin{equation}
V=8 g^2 \left[-3-2 \cosh(\sqrt{2} b_2)+\cosh(\sqrt{2} b_1)
(2+\cosh(\sqrt{2} b_2))\right].
\end{equation}
There is a $(0,6)$ critical point at $b_1=b_2=0$ with $V_0=-16g^2$,
and the associated superconformal algebra is given by $SU(1,1|3)$.
The scalar mass spectrum is given in the table below.
\begin{center}
\begin{tabular}{|c|c|c|}
  \hline
  $m^2L^2$ &  $(h,\bar{h})$ & $SU(3)$  \\ \hline
   $-1$& $\left(\frac{1}{2},\frac{1}{2}\right)$ & $\mathbf{3}+\bar{\mathbf{3}}$ \\
  \hline
  $3$& $\left(\frac{3}{2},\frac{3}{2}\right)$ & $2\times\mathbf{1}$ \\
  \hline
\end{tabular}
\end{center}

\subsubsection{$SU(2,1)\times SU(2,1)$ gauging}
There are eight scalars parametrized by $\frac{SU(2,1)}{SU(2)\times
U(1)}\times \frac{SU(2,1)}{SU(2)\times U(1)}$ where the two
$SU(2,1)\subset SU(4,2)$ are different subgroups from those
appearing in the gauge group. The coset representative is given by
\begin{equation}
L= e^{a_1q_1}e^{a_2q_2}e^{a_3q_3}e^{\frac{i}{\sqrt
{2}}b_1c_{16}}e^{a_4\tilde{w}_1}e^{a_5\tilde{w}_2}e^{a_6\tilde{w}_3}e^{\frac{i}{\sqrt{2}}b_2c_{29}},
\end{equation}
where $q_i$'s are given in \eqref{qi_def} and
\begin{equation}
\tilde{w}_1=\frac{1}{2}c_{13},\qquad
\tilde{w}_2=\frac{1}{2}c_{14},\qquad
\tilde{w}_3=\frac{1}{2}\left(-\frac{1}{\sqrt{3}}c_8+\frac{2}{\sqrt{6}}c_{15}\right).
\end{equation}
We find the potential
\begin{eqnarray}
V&=&2 g^2 \left[\cosh[\sqrt{2} (b_1+b_2)]-\sinh[\sqrt{2} (b_1+b_2)] (1+\cosh(2 \sqrt{2} b_1)+\cosh(2 \sqrt{2} b_2)\right.\nonumber\\
&&\left.-4 \cosh(\sqrt{2} (b_1+b_2))+\cosh(2 \sqrt{2} (b_1+b_2))+\sinh(2 \sqrt{2} b_1)+\sinh(2 \sqrt{2} b_2)\right.\nonumber\\
&&\left.-4 \sinh(\sqrt{2} (b_1+b_2))+\sinh(2 \sqrt{2}
(b_1+b_2))\right].
\end{eqnarray}
There is no non-trivial critical point. The trivial one is given by
$b_1=b_2=0$ with $V_0=0$ and preserves $N=6$ supersymmetry. The
scalar mass spectrum is given in the table below.
\begin{center}
\begin{tabular}{|c|c|c|}
  \hline
  $m^2$ &   $SU(2)\times SU(2)$  \\ \hline
   $16g^2$&  $2\times(\mathbf{2},\mathbf{1})+2\times(\mathbf{1},\mathbf{2})$ \\
  \hline
\end{tabular}
\end{center}

\subsubsection{$SU(4,1)\times U(1)$ gauging}
In this case, there are eight scalars parametrized by
$\frac{SU(4,1)}{SU(4)\times U(1)}$ whose coset representative is
given by
\begin{equation}
L=
e^{a_1c_3}e^{a_2c_2}e^{a_3c_3}e^{a_4c_5}e^{\frac{1}{\sqrt{3}}a_5c_8}e^{a_6c_{10}}e^{\frac{1}{\sqrt{6}}a_7c_{15}}e^{\frac{i}{\sqrt{2}}bc_{26}}.
\end{equation}
The potential is obtained to be
\begin{equation}
V=-8 g^2 \left(5+3 \cosh(\sqrt{2} b)\right)
\end{equation}
which does not admit any non-trivial critical points. At $b=0$, we
find $V_0=-64g^2$ and $(0,6)$ supersymmetry. In this case, the full
R-symmetry $SU(4)\subset SU(4,1)$ is preserved, and the
corresponding superconformal algebra is given by
$OSp(6|2,\mathbb{R})$. The scalar mass spectrum is given in the
table below.
\begin{center}
\begin{tabular}{|c|c|c|}
  \hline
  $m^2L^2$ &  $(h,\bar{h})$ & $SU(4)$  \\ \hline
   $-\frac{3}{4}$& $\left(\frac{1}{4},\frac{1}{4}\right)$ & $\mathbf{4}$ \\
  \hline
  $-\frac{3}{4}$& $\left(\frac{3}{4},\frac{3}{4}\right)$ & $\bar{\mathbf{4}}$ \\
  \hline
\end{tabular}
\end{center}

\subsection{The $k=3$ case}
In this case, we find the following gauge groups:
\begin{eqnarray}
SU(3,3)\times U(1)&:&\qquad \Theta=\Theta_{SU(3,3)} - \frac{5}{7}
\Theta_{U(1)}\\
SU(2,3)\times SU(2) \times U(1)&:&\qquad
\Theta=\Theta_{SU(2,3)}-\Theta_{SU(2)} - \frac{3}{7} \Theta_{U(1)}\\
SU(1,3)\times SU(3)\times U(1)&:&\qquad
\Theta=\Theta_{SU(1,3)}-\Theta_{SU(3)} - \frac{1}{7} \Theta_{U(1)}\\
SU(3,2)\times SU(1,1) \times U(1)&:&\qquad \Theta=\Theta_{SU(3,2)}
-\Theta_{SU(1,1)}- \frac{3}{7} \Theta_{U(1)}\\
SU(2,2)\times SU(2,1)\times U(1)&:&\qquad
\Theta=\Theta_{SU(2,2)}-\Theta_{SU(2,1)} -\frac{1}{7}\Theta_{U(1)}\\
SU(1,2)\times SU(3,1)\times U(1)&:&\qquad \Theta=\Theta_{SU(1,2)} -
\Theta_{SU(3,1)} +\frac{1}{7}\Theta_{U(1)}\\
SU(4,1)\times SU(2)\times U(1)&:&\qquad \Theta=\Theta_{SU(4,1)} -
\Theta_{SU(2)} -\frac{3}{7}\Theta_{U(1)}\\
SU(4,2)\times U(1)&:&\qquad \Theta=\Theta_{SU(4,2)}
-\frac{5}{7}\Theta_{U(1)}\, .
\end{eqnarray}
We now study their critical points.
\subsubsection{$SU(3,3)\times U(1)$ gauging}
There are six scalars parametrizing the coset
$\frac{SU(3,1)}{SU(3)\times U(1)}$. The coset representative takes
the form
\begin{equation}
L=
e^{a_1L_3}e^{a_2L_2}e^{a_3L_3}e^{a_4L_5}e^{\frac{1}{\sqrt{3}}a_5L_8}e^{\frac{i}{\sqrt{2}}b
c_{17}},
\end{equation}
where
\begin{eqnarray}
L_2&=& \frac{1}{2}c_{34},\qquad L_3=
\frac{1}{10}\left(\sqrt{10}c_{24}-\sqrt{15}c_{35}\right),\nonumber
\\
L_5&=&\frac{1}{2}c_{45},\qquad L_8=
-\frac{1}{40}\left(2\sqrt{15}c_{24}+2\sqrt{10}c_{35}-5\sqrt{14}c_{48}\right).\label{Li_def}
\end{eqnarray}
The potential is given by
\begin{equation}
V=8g^2(3\cosh(\sqrt{2}b)-5)
\end{equation}
which admits only a $(0,6)$ trivial critical point at $b=0$ with
$V_0=-16g^2$, and the associated superconformal algebra is given by
$SU(1,1|3)$. The scalar mass spectrum is given in the table below.
\begin{center}
\begin{tabular}{|c|c|c|}
  \hline
  $m^2L^2$ &  $(h,\bar{h})$ & $SU(3)\times SU(3)$  \\ \hline
   $3$& $\left(\frac{3}{2},\frac{3}{2}\right)$ & $(\mathbf{3},\mathbf{1})+(\bar{\mathbf{3}},\mathbf{1})$ \\
  \hline
\end{tabular}
\end{center}

\subsubsection{$SU(2,3)\times SU(2) \times U(1)$ gauging}
The twelve scalars parametrizing the coset
$\frac{SU(2,3)}{SU(2)\times SU(3)\times U(1)}$ whose coset
representative is given by
\begin{equation}
L=
e^{a_1L_3}e^{a_2L_2}e^{a_3L_3}e^{a_4L_5}e^{\frac{1}{\sqrt{3}}a_5L_8}
e^{a_6L_3}e^{a_7L_2}e^{a_8L_3}e^{a_9c_3}e^{a_{10}c_2}
e^{\frac{i}{2}b_1c_{16}}e^{\frac{i}{2}b_2c_{27}},
 \end{equation}
where $L_i$'s are given in \eqref{Qi_def}. We find the potential
\begin{equation}
V=-8g^2[3+\cosh(\sqrt{2}b_1)(-2+\cosh(\sqrt{2}b_2))-2\cosh(\sqrt{2}b_2)]\,
.
\end{equation}
There are two critical point:
\begin{itemize}
\item $b_1=b_2=0$ with $V_0=0$ and $N=6$ supersymmetry
\item $b_1=b_2=\frac{1}{\sqrt{2}}\cosh^{-1}{2}$ with $V_0=8g^2$ and residual symmetry $SU(2)_{\textrm{diag}}\times SU(2)\times U(1)\times U(1)$.
The $SU(2)_{\textrm{diag}}$ here is a diagonal subgroup of the
$SU(2)$ and the $SU(2)$ in the $SU(3)\subset SU(2,3)$.
\end{itemize}
The scalar mass spectrum at the trivial critical point is given in
the table below.
\begin{center}
\begin{tabular}{|c|c|}
  \hline
  $m^2$ &   $SU(3)\times SU(2)\times SU(2)$  \\ \hline
   $16g^2$&  $(\mathbf{3},\mathbf{1},\mathbf{2})+(\bar{\mathbf{3}},\mathbf{1},\mathbf{2})$ \\
  \hline
\end{tabular}
\end{center}

\subsubsection{$SU(1,3)\times SU(3)$ gauging}
The eighteen scalars parametrizing $\frac{SU(3,3)}{SU(3)\times
SU(3)\times U(1)}$ coset are encoded in the coset representative
\begin{eqnarray}
L&=&
e^{a_1q_3}e^{a_2q_2}e^{a_3q_3}e^{a_4q_5}e^{\frac{1}{\sqrt{3}}a_5q_8}e^{a_6q_3}
e^{a_7q_2}e^{a_8q_3}e^{a_9L_3}e^{a_{10}L_2}e^{a_{11}L_3}e^{a_{12}L_5}
e^{\frac{1}{\sqrt{3}}a_{13}L_8}e^{a_{14}L_3}\times \nonumber\\
&
&e^{a_{15}L_2}e^{\frac{i}{\sqrt{2}}b_1c_{16}}e^{\frac{i}{\sqrt{2}}b_2c_{27}}e^{\frac{i}{\sqrt{2}}b_3c_{40}},
\end{eqnarray}
where $q_i$'s and $L_i$'s are the same as those given in
\eqref{qi_def} and \eqref{Li_def}, respectively. The potential is
given by
\begin{eqnarray}
V&=&-4 g^2\left[\frac{3 }{2}-\frac{1}{2} \cosh(2 \sqrt{2}
b_1)-\frac{1}{2}  \cosh(2 \sqrt{2} b_2)+\frac{1}{4}
\left(-2+2 \cosh(\sqrt{2} b_1)\right .\right .\nonumber \\
& &\left . \left .+2 \cosh(\sqrt{2} b_2)+2 \cosh(\sqrt{2}
b_3)\right)^2-\frac{1}{2} \cosh(2 \sqrt{2} b_3)\right].
\end{eqnarray}
The only one $(0,6)$ critical point is given by $b_1=b_2=b_3=0$ with
$V_0=-16g^2$, and the associated superconformal algebra is given by
$SU(1,1|3)$. The scalar mass spectrum is given in the table below.
\begin{center}
\begin{tabular}{|c|c|c|}
  \hline
  $m^2L^2$ &  $(h,\bar{h})$ & $SU(3)\times SU(3)$  \\ \hline
   $-1$& $\left(\frac{1}{2},\frac{1}{2}\right)$ & $(\mathbf{3},\bar{\mathbf{3}})+(\bar{\mathbf{3}},\mathbf{3})$ \\
  \hline
\end{tabular}
\end{center}

\subsubsection{$SU(3,2)\times SU(1,1) \times U(1)$ gauging}
The ten relevant scalars are described by the coset representative
of the coset $\frac{SU(2,1)}{SU(2)\times U(1)}\times
\frac{SU(3,1)}{SU(3)\times U(1)}$. We choose the parametrization of
$L$ to be
\begin{equation}
   L= e^{a_1c_{33}}e^{a_2c_{34}}e^{\frac{1}{\sqrt{2}}a_3\left[c_{33},c_{34}\right]}e^{\frac{i}{\sqrt{2}}b_1 c_{16}}
   e^{a_4w_3}e^{a_5w_2}e^{a_6w_3}e^{a_7w_5}
   e^{\frac{1}{\sqrt{3}}\frac{a_8}{\sqrt{3}}w_8}e^{\frac{i}{\sqrt{2}}b_2c_{39}},
\end{equation}
where $w_i$'s are given in \eqref{wi_def}. The potential is found to
be
\begin{eqnarray}
V&=&2 g^2 \left[\cosh(\sqrt{2} (b_1+b_2))-\sinh(\sqrt{2} (b_1+b_2))) (1-4 \cosh(\sqrt{2} b_1)+\cosh(2 \sqrt{2} b_1)\right.\nonumber\\
& &+4 \cosh(\sqrt{2} b_2)\left.+\cosh(2 \sqrt{2} b_2)-12
\cosh(\sqrt{2} (b_1+b_2))+\cosh(2 \sqrt{2} (b_1+b_2))\right.\nonumber\\
&&\left. +4 \cosh(\sqrt{2} (2 b_1+b_2))-4 \cosh(\sqrt{2} (b_1+2
b_2)) -4 \sinh(\sqrt{2} b_1)+\sinh(2 \sqrt{2} b_1)
\right.\nonumber\\
&&\left.+4 \sinh(\sqrt{2} b_2)+\sinh(2 \sqrt{2} b_2)-12
\sinh(\sqrt{2} (b_1+b_2))+\sinh(2 \sqrt{2} (b_1+b_2))
\right.\nonumber\\
&&\left.+4 \sinh(\sqrt{2} (2 b_1+b_2))-4 \sinh(\sqrt{2} (b_1+2
b_2))\right].
\end{eqnarray}
There is a trivial $(0,6)$ critical point at $b_1=b_2=0$ with
$V_0=-16g^2$, and the associated superconformal algebra is given by
$SU(1,1|3)$. The scalar mass spectrum is given in the table below.
\begin{center}
\begin{tabular}{|c|c|c|}
  \hline
  $m^2L^2$ &  $(h,\bar{h})$ & $SU(3)\times SU(2)$  \\ \hline
   $3$& $\left(\frac{3}{2},\frac{3}{2}\right)$ & $2\times(\mathbf{1},\mathbf{2})$ \\
  \hline
  $-1$& $\left(\frac{1}{2},\frac{1}{2}\right)$ & $(\mathbf{3},\mathbf{1})+(\bar{\mathbf{3}},\mathbf{1})$ \\
  \hline
\end{tabular}
\end{center}

\subsubsection{$SU(2,2)\times SU(2,1)\times U(1)$ gauging}
The twelve scalars are parametrized by the following coset
representative
\begin{equation}
  L= e^{a_1q_1}e^{a_2q_2} e^{a_3q_3}e^{a_4Q_1} e^{a_5Q_2}e^{a_6Q_3} e^{\frac{i}{\sqrt{2}}b_1c_{16}}e^{\frac{i}{\sqrt{2}}b_2c_{27}}
  e^{a_7Z}e^{a_8c_{14}} e^{a_9Z}e^{\frac{i}{\sqrt{2}}b_3c_{40}},
\end{equation}
where $q_{i}=\frac{1}{2}c_{i}$,
\begin{equation}
Z=\frac{1}{\sqrt{2}}[c_{13},c_{14}],
\end{equation}
and $Q_i$'s are given in \eqref{Qi_def}. This $L$ describes the
coset $\frac{SU(2,2)}{SU(2)\times SU(2)\times U(1)}\times
\frac{SU(2,1)}{SU(2)\times U(1)}$. The potential is given by
\begin{eqnarray}
V&=&-g^2 \left[6-2 \cosh(2 \sqrt{2} b_1)-2 \cosh(2 \sqrt{2} b_2)-2
\cosh(2 \sqrt{2} b_3)\right.\nonumber\\
&&\left.+4 \left(-1+\cosh(\sqrt{2} b_1) +\cosh(\sqrt{2}
b_2)-\cosh(\sqrt{2} b_3)\right)^2\right].
\end{eqnarray}
There two critical points:
\begin{itemize}
\item at $b_1=b_2=b_3=0$ with $V_0=0$ and $N=6$ supersymmetry
\item at $b_1=b_2=\frac{1}{\sqrt{2}}\cosh^{-1}{2}$, $b_3=0$ with $V_0=8g^2$ and residual symmetry $SU(2)_{\textrm{diag}}\times
SU(2)\times U(1)\times U(1)$. The $SU(2)_{\textrm{diag}}$ is a
diagonal subgroup of the $SU(2)\subset SU(2,1)$ and one of the
$SU(2)\subset SU(2,2)$.
\end{itemize}
The scalar mass spectrum at the trivial critical point is given in
the table below.
\begin{center}
\begin{tabular}{|c|c|}
  \hline
  $m^2$ &   $SU(2)\times SU(2)\times SU(2)$  \\ \hline
   $16g^2$&  $2\times (\mathbf{1},\mathbf{2},\mathbf{1})+2\times(\mathbf{2},\mathbf{1},\mathbf{2})$ \\
  \hline
\end{tabular}
\end{center}

\subsubsection{$SU(1,2)\times SU(3,1)\times U(1)$ gauging}
The fourteen scalars describing the coset
$\frac{SU(1,1)}{U(1)}\times \frac{SU(3,2)}{SU(3)\times SU(2)\times
U(1)}$ are parametrized by
\begin{eqnarray}
L&=& e^{a_1c_3} e^{a_2c_2}e^{a_3c_3}e^{a_4c_5}
e^{\frac{1}{\sqrt{3}}a_5c_8} e^{a_6c_3}e^{a_7c_2}e^{a_8c_3}
e^{a_9Q_3}e^{a_{10}Q_2}e^{\frac{i}{\sqrt{2}}b_1c_{16}}e^{\frac{i}{\sqrt{2}}b_2c_{27}}\times
\nonumber \\ & &
e^{\frac{1}{\sqrt{2}}a_{11}\left[c_{23},c_{24}\right]}e^{\frac{i}{\sqrt{2}}b_3c_{42}},
\end{eqnarray}
where $Q_i$'s are given in \eqref{Qi_def}. The potential is given by
\begin{eqnarray}
V&=&-g^2 \left[6-2 \cosh(2 \sqrt{2} b_1)-2 \cosh(2 \sqrt{2} b_2)-2
\cosh(2 \sqrt{2} b_3)\right.\nonumber\\
&&\left.+4 \left(1+\cosh(\sqrt{2} b_1) +\cosh(\sqrt{2}
b_2)-\cosh(\sqrt{2} b_3)\right)^2\right].
\end{eqnarray}
The trivial $(0,6)$ critical point at $b_1=b_2=b_3=0$ has
$V_0=-16g^2$, and the associated superconformal algebra is given by
$SU(1,1|3)$. The scalar mass spectrum is given in the table below.
\begin{center}
\begin{tabular}{|c|c|c|}
  \hline
  $m^2L^2$ &  $(h,\bar{h})$ & $SU(3)\times SU(2)$  \\ \hline
  $-1$& $\left(\frac{1}{2},\frac{1}{2}\right)$ & $(\mathbf{3},\mathbf{2})+(\bar{\mathbf{3}},\mathbf{2})+2\times (\mathbf{1},\mathbf{1})$ \\
  \hline
\end{tabular}
\end{center}

\subsubsection{$SU(4,1)\times SU(2)\times U(1)$ gauging}
The sixteen scalars parametrize the coset
$\frac{SU(4,2)}{SU(4)\times SU(2)\times U(1)}$ whose coset
representative takes the form
\begin{eqnarray}
L&=& e^{a_1q_3} e^{a_2q_2}e^{a_3q_3}e^{a_4q_5}
e^{\frac{1}{\sqrt{3}}a_5q_8} e^{a_6q_{10}}e^{a_7q_3}e^{a_8q_2}
e^{a_9q_3}e^{a_{10}q_5}e^{\frac{1}{\sqrt{3}}a_{11}q_8}e^{a_{12}\tilde{q}_3}
e^{a_{13}\tilde{q}_2}\times
\nonumber\\
&
&e^{a_{14}\tilde{q}_3}e^{\frac{i}{\sqrt{2}}b_1c_{25}}e^{\frac{i}{\sqrt{2}}b_2c_{38}}
\end{eqnarray}
where $q_{i}=\frac{1}{2}c_{i}$ and
\begin{equation}
\tilde{q}_2=\frac{1}{2}c_{47},\qquad
\tilde{q}_3=-\frac{1}{12}\left(\sqrt{15}c_{35}-\sqrt{21}c_{48}\right).
\end{equation}
We find the corresponding potential
\begin{equation}
V=-8 g^2 \left(3+2 \cosh(\sqrt{2} b_2)+\cosh(\sqrt{2} b_1)
(2+\cosh(\sqrt{2} b_2))\right).
\end{equation}
The only one $(0,6)$ critical point is given by $b_1=b_2=0$ with
$V_0=-64g^2$. The superconformal algebra in this case is given by
$OSp(6|2,\mathbb{R})$ since the full $SU(4)$ R-symmetry is
preserved. The scalar mass spectrum is given in the table below.
\begin{center}
\begin{tabular}{|c|c|c|}
  \hline
  $m^2L^2$ &  $(h,\bar{h})$ & $SU(4)\times SU(2)$  \\ \hline
  $-\frac{3}{4}$& $\left(\frac{1}{4},\frac{1}{4}\right)$ & $(\mathbf{4},\mathbf{2})$ \\
  \hline
  $-\frac{3}{4}$& $\left(\frac{3}{4},\frac{3}{4}\right)$ & $(\bar{\mathbf{4}},\mathbf{2})$ \\
  \hline
\end{tabular}
\end{center}

\subsubsection{$SU(4,2)\times U(1)$ gauging}
The coset representative for the eight scalars parametrizing the
coset $\frac{SU(4,1)}{SU(4)\times U(1)}$ takes the form
\begin{equation}
L= e^{a_1q_3} e^{a_2q_2}e^{a_3q_3}e^{a_4q_5}
e^{\frac{1}{\sqrt{3}}a_5q_8}
e^{a_6q_{10}}e^{\frac{1}{\sqrt{6}}a_7q_{15}}e^{\frac{i}{\sqrt{2}}b
c_{37}},
\end{equation}
where $q_i$'s are given in \eqref{qi_def}. We find the following
potential
\begin{equation}
V=-8 g^2 \left(5+3 \cosh(\sqrt{2} b)\right)
\end{equation}
which does not admit any non-trivial critical points. The trivial
$(0,6)$ critical point is at $b=0$ with $V_0=-64g^2$. The
superconformal symmetry is again $OSp(6|2,\mathbb{R})$. The scalar
mass spectrum is given in the table below.
\begin{center}
\begin{tabular}{|c|c|c|}
  \hline
  $m^2L^2$ &  $(h,\bar{h})$ & $SU(4)\times SU(2)$  \\ \hline
  $-\frac{3}{4}$& $\left(\frac{1}{4},\frac{1}{4}\right)$ & $(\mathbf{4},\mathbf{1})$ \\
  \hline
  $-\frac{3}{4}$& $\left(\frac{3}{4},\frac{3}{4}\right)$ & $(\bar{\mathbf{4}},\mathbf{1})$ \\
  \hline
\end{tabular}
\end{center}

\subsection{The $k=4$ case}
In this case, we find the following gauge groups:
\begin{eqnarray}
SU(3,4)\times U(1)&:&\qquad \Theta=\Theta_{SU(3,4)} - \frac{3}{4}
\Theta_{U(1)}\\
SU(2,4)\times SU(2) \times U(1)&:&\qquad
\Theta=\Theta_{SU(2,4)}-\Theta_{SU(2)} -
\frac{1}{2} \Theta_{U(1)}\\
SU(1,4)\times SU(3)\times U(1)&:&\qquad
\Theta=\Theta_{SU(1,4)}-\Theta_{SU(3)} - \frac{1}{4} \Theta_{U(1)}\\
SU(3,3)\times SU(1,1) \times U(1)&:&\qquad \Theta=\Theta_{SU(3,3)}
-\Theta_{SU(1,1)}- \frac{1}{2} \Theta_{U(1)}\\
SU(2,3)\times SU(2,1)\times U(1)&:&\qquad
\Theta=\Theta_{SU(2,3)}-\Theta_{SU(2,1)} -
\frac{1}{4}\Theta_{U(1)}\\
SU(1,3)\times SU(3,1)&:&\qquad \Theta=\Theta_{SU(1,3)} -
\Theta_{SU(3,1)}\\
SU(2,2)\times SU(2,2)&:&\qquad \Theta=\Theta_{SU(2,2)} -
\Theta_{SU(2,2)}\, .
\end{eqnarray}
We then move to the study of their critical points.
\subsubsection{$SU(3,4)\times U(1)$ gauging}
There are eight scalars described by the coset
$\frac{SU(4,1)}{SU(4)\times U(1)}$ whose coset representative is
given by
\begin{equation}
L= e^{a_1j_3} e^{a_2j_2}e^{a_3j_3}e^{a_4j_5}
e^{\frac{1}{\sqrt{3}}a_5j_8}
e^{a_6j_{10}}e^{\frac{1}{\sqrt{6}}a_7j_{15}}e^{\frac{i}{\sqrt{2}}bc_{17}},
\end{equation}
where
\begin{eqnarray}
j_2 &=& \frac{1}{2}c_{34},\qquad
j_3=\frac{1}{2\sqrt{2}}\left[c_{33},c_{34}\right],\qquad j_5 =
\frac{1}{2}c_{45},\nonumber \\
j_8&=&\frac{1}{15}(3\sqrt{10}c_{24}+2\sqrt{15}c_{35}-5\sqrt{21}c_{48}),
\qquad j_{10} = \frac{1}{2}c_{58},\nonumber \\
j_{15}&=&\frac{1}{105}(21\sqrt{10}c_{24}+14\sqrt{15}c_{35}+10\sqrt{21}c_{48}-90\sqrt{7}c_{63}).\label{ji_def}
\end{eqnarray}
We find the potential
\begin{equation}
V=8 g^2 \left(-5+3 \cosh(\sqrt{2} b)\right)
\end{equation}
whose only one $(0,6)$ critical point is characterized by $b=0$ and
$V_0=-16g^2$, and the associated superconformal algebra is given by
$OSp(6|2,\mathbb{R})$. The scalar mass spectrum is given in the
table below.
\begin{center}
\begin{tabular}{|c|c|c|}
  \hline
  $m^2L^2$ &  $(h,\bar{h})$ & $SU(4)\times SU(3)$  \\ \hline
  $3$& $\left(\frac{3}{2},\frac{3}{2}\right)$ & $(\mathbf{4},\mathbf{1})+(\bar{\mathbf{4}},\mathbf{1})$ \\
  \hline
\end{tabular}
\end{center}

\subsubsection{$SU(2,4)\times SU(2) \times U(1)$ gauging}
The sixteen scalars are described by the coset
$\frac{SU(4,2)}{SU(4)\times SU(2)\times U(1)}$. The coset
representative takes the form
\begin{eqnarray}
L&=& e^{a_1j_3}e^{a_2j_2} e^{a_3 j_3}e^{a_4 j_5}
e^{\frac{1}{\sqrt{3}}a_5j_8}e^{a_6j_3}
e^{a_7j_3}e^{a_8j_2}e^{a_9j_3} e^{a_{10} j_5 }
e^{\frac{1}{\sqrt{3}}a_{11} j_8}e^{a_{12} c_3 }e^{a_{13} c_2} \times
\nonumber \\ & &e^{a_{14} c_3
}e^{\frac{i}{\sqrt{2}}b_1c_{16}}e^{\frac{i}{\sqrt{2}}b_2c_{27}},
\end{eqnarray}
where $j_i$'s are given in \eqref{ji_def}. The corresponding
potential is given by
\begin{equation}
V=-8 g^2 \left[3+\cosh(\sqrt{2} b_1) (-2+\cosh(\sqrt{2} b_2))-2
\cosh(\sqrt{2} b_2)\right].
\end{equation}
There are two critical points:
\begin{itemize}
\item at $b_1=b_2=0$ with $V_0=0$ and $N=6$ supersymmetry
\item at $b_1=b_2=\frac{1}{\sqrt{2}}\cosh^{-1} {2}$ with $V_0=8g^2$ and residual symmetry $SU(2)_{\textrm{diag}}\times
SU(2)\times SU(2)\times U(1)\times U(1)$. The
$SU(2)_{\textrm{diag}}$ is a diagonal subgroup of the $SU(2)$ and an
$SU(2)$ in the $SU(4)\subset SU(2,4)$.
\end{itemize}
The scalar mass spectrum at the trivial critical point is given in
the table below.
\begin{center}
\begin{tabular}{|c|c|}
  \hline
  $m^2$ &   $SU(4)\times SU(2)\times SU(2)$  \\ \hline
   $16g^2$&  $(\mathbf{4},\mathbf{1},\mathbf{2})+(\bar{\mathbf{2}},\mathbf{1},\mathbf{2})$ \\
  \hline
\end{tabular}
\end{center}

\subsubsection{$SU(1,4)\times SU(3)\times U(1)$ gauging}
All twenty four scalars are encoded in the coset
$\frac{SU(4,3)}{SU(4)\times SU(3)\times U(1)}$. The coset
representative takes the form
\begin{eqnarray}
L&=& e^{a_1c_3}e^{a_2c_2} e^{a_3 c_3}e^{a_4 c_5}
e^{\frac{1}{\sqrt{3}}a_5c_8}e^{a_6c_3}
e^{a_7c_3}e^{a_8j_3}e^{a_9j_2}
e^{a_{10} j_3 } e^{a_{11} j_5 }e^{\frac{1}{\sqrt{3}}a_{12} j_8}e^{a_{13} j_{10} }e^{a_{14} j_3 }\times \nonumber\\
& &e^{a_{14} j_3 }e^{a_{15} j_2 }e^{a_{16} j_3 }e^{a_{17} j_5
}e^{\frac{1}{\sqrt{3}}a_{18} j_8}e^{a_{19} j_3 }e^{a_{20} j_2
}e^{a_{21} j_3
}e^{\frac{i}{\sqrt{2}}b_1c_{16}}e^{\frac{i}{\sqrt{2}}b_2c_{27}}e^{\frac{i}{\sqrt{2}}b_3c_{40}},
\end{eqnarray}
where $j_i$'s are given in \eqref{ji_def}. We find the potential
\begin{eqnarray}
V&=&-8 g^2 \left[2+\cosh(\sqrt{2} b_2) (-1+\cosh(\sqrt{2} b_3))-\cosh(\sqrt{2} b_3)+\cosh(\sqrt{2} b_1)\times  \right.\nonumber\\
&&\left.(-1+\cosh(\sqrt{2} b_2)+\cosh(\sqrt{2} b_3))\right].
\end{eqnarray}
The only one $(0,6)$ critical point is at $b_1=b_2=b_3=0$ with
$V_0=-16g^2$. The superconformal symmetry is $OSp(6|2,\mathbb{R})$.
The scalar mass spectrum is given in the table below.
\begin{center}
\begin{tabular}{|c|c|c|}
  \hline
  $m^2L^2$ &  $(h,\bar{h})$ & $SU(4)\times SU(3)$  \\ \hline
  $-1$& $\left(\frac{1}{2},\frac{1}{2}\right)$ & $(\mathbf{4},\bar{\mathbf{3}})+(\bar{\mathbf{4}},\mathbf{3})$ \\
  \hline
\end{tabular}
\end{center}

\subsubsection{$SU(3,3)\times SU(1,1) \times U(1)$ gauging}
In this case, there are twelve scalars described by the coset
$\frac{SU(3,1)}{SU(3)\times U(1)}\times \frac{SU(3,1)}{SU(3)\times
U(1)}$. The coset representative takes the form
\begin{equation}
L= e^{a_1w_3} e^{a_2w_2}e^{a_3w_3}e^{a_4w_5}
e^{\frac{1}{\sqrt{3}}a_5w_8}e^{\frac{i}{\sqrt{2}}b_1c_{52}}
e^{a_6L_3}e^{a_7L_2}e^{a_8L_3}
e^{a_9L_5}e^{\frac{1}{\sqrt{3}}a_{10}L_8}e^{\frac{i}{\sqrt{2}}b_2c_{17}},
\end{equation}
where $w_i$'s and $L_i$'s are given in \eqref{wi_def} and
\eqref{Li_def}, respectively. We find the potential
\begin{equation}
V=8 g^2 [-3+\cosh(\sqrt{2} b_1) (-2+\cosh(\sqrt{2} b_2))+2
\cosh(\sqrt{2} b_2)]
\end{equation}
whose trivial $(0,6)$ critical point is characterized by $b_1=b_2=0$
with $V_0=-16g^2$, and the associated superconformal algebra is
given by $SU(1,1|3)$. The scalar mass spectrum is given in the table
below.
\begin{center}
\begin{tabular}{|c|c|c|}
  \hline
  $m^2L^2$ &  $(h,\bar{h})$ & $SU(3)\times SU(3)$  \\ \hline
  $3$& $\left(\frac{3}{2},\frac{3}{2}\right)$ & $(\mathbf{1},\mathbf{3})+(\mathbf{1},\bar{\mathbf{3}})$ \\
  \hline
  $-1$& $\left(\frac{1}{2},\frac{1}{2}\right)$ & $(\mathbf{3},\mathbf{1})+(\bar{\mathbf{3}},\mathbf{1})$ \\
  \hline
\end{tabular}
\end{center}

\subsubsection{$SU(2,3)\times SU(2,1)\times U(1)$ gauging}
The sixteen scalars described by the coset
$\frac{SU(2,1)}{SU(2)\times U(1)}\times \frac{SU(3,2)}{SU(3)\times
SU(2)\times U(1)}$ are parametrized by the coset representative
\begin{eqnarray}
L&=&
e^{a_1L_3}e^{a_2L_2}e^{a_3L_3}e^{a_4L_5}e^{\frac{1}{\sqrt{3}}a_5L_8}
e^{a_6L_3}e^{a_7L_2}e^{a_8L_3}e^{a_9q_3}e^{a_{10}q_2}
e^{\frac{i}{\sqrt{2}}b_2c_{16}}e^{\frac{i}{\sqrt{2}}b_3c_{27}}\times
\nonumber
\\ & &e^{a_{11}z_1}e^{a_{12}z_2}e^{a_{13}z_3}e^{\frac{i}{\sqrt{2}}b_1c_{44}},
\end{eqnarray}
where
\begin{equation}
z_1=\frac{1}{2}c_{13},\qquad z_2=\frac{1}{2}c_{14}, \qquad
z_3=\frac{1}{2}\left(\frac{-1}{\sqrt{3}}c_8+\frac{2}{\sqrt{6}}c_{15}\right),
\label{zi_def}
\end{equation}
and $q_i$'s and $L_i$'s are given in \eqref{qi_def} and
\eqref{Li_def}, respectively. We find the potential
\begin{eqnarray}
V&=&8 g^2 \left[-2+\cosh(\sqrt{2} b_3)+\cosh(\sqrt{2} b_1) (-1+\cosh(\sqrt{2} b_2)+\cosh(\sqrt{2} b_3))\right.\nonumber\\
&&\left.-2 \cosh(\sqrt{2} b_2)
\sinh^2\left(\frac{b_3}{\sqrt{2}}\right)\right].
\end{eqnarray}
There are two critical points:
\begin{itemize}
\item at $b_1=b_2=b_3=0$ with $V_0=0$ and $N=6$ supersymmetry
\item at $b_1=0$, $b_2=b_3=\frac{1}{\sqrt{2}}\cosh^{-1}{2}$ with $V_0=8g^2$ and residual symmetry $SU(2)_{\textrm{diag}}\times
SU(2)\times U(1)\times U(1)\times U(1)$. The $SU(2)_{\textrm{diag}}$
is a diagonal subgroup of the $SU(2)\subset SU(2,1)$ and the $SU(2)$
subgroup of the $SU(3)\subset SU(2,3)$.
\end{itemize}
The scalar mass spectrum at the trivial critical point is given in
the table below.
\begin{center}
\begin{tabular}{|c|c|}
  \hline
  $m^2$ &   $SU(3)\times SU(2)\times SU(2)$  \\ \hline
   $16g^2$&  $2\times (\mathbf{1},\mathbf{2},\mathbf{1})+(\mathbf{2},\mathbf{1},\mathbf{3})+(\mathbf{2},\mathbf{1},\bar{\mathbf{3}})$ \\
  \hline
\end{tabular}
\end{center}

\subsubsection{$SU(1,3)\times SU(3,1)$ gauging}
The twenty scalars parametrizing the coset
$\frac{SU(1,1)}{U(1)}\times \frac{SU(3,3)}{SU(3)\times SU(3)\times
U(1)}$ are encoded in the following coset representative
\begin{eqnarray}
L&=&
e^{\frac{1}{\sqrt{2}}a_1\left[c_{31},c_{32}\right]}e^{\frac{i}{\sqrt{2}}b_1c_{56}}
e^{a_2 c_3}e^{a_4 c_3} e^{a_5c_5}e^{\frac{1}{\sqrt{3}}a_6c_8}
e^{a_7c_3}e^{a_8c_2}e^{a_9c_3} e^{a_{10} L_3 } e^{a_{11} L_2 }e^{a_{12} L_3}e^{a_{13} L_5 }\times \nonumber\\
&&e^{\frac{1}{\sqrt{3}}a_{14} L_8}e^{a_{15} L_3 }e^{a_{16}
L_2}e^{\frac{i}{\sqrt{2}}b_2c_{16}}e^{\frac{i}{\sqrt{2}}b_3c_{27}}e^{\frac{i}{\sqrt{2}}b_4c_{40}},
\end{eqnarray}
where $L_i$'s are given in \eqref{Li_def}. We find the potential
\begin{eqnarray}
V&=&-g^2 \left[6-2 \cosh(2 \sqrt{2} b_2)-2 \cosh(2 \sqrt{2} b_3) +4
\left(-\cosh(\sqrt{2} b_1)+\cosh(\sqrt{2} b_2)\right.\right.\nonumber\\
&&\left.\left.+\cosh(\sqrt{2} b_3)+\cosh(\sqrt{2} b_4)\right)^2-2
\cosh(2 \sqrt{2} b_4)-4 \sinh^2(\sqrt{2} b_1)\right]
\end{eqnarray}
which admits only a trivial $(0,6)$ critical point at
$b_1=b_2=b_3=0$ with $V_0=-16g^2$, and the associated superconformal
algebra is given by $SU(1,1|3)$. The scalar mass spectrum is given
in the table below.
\begin{center}
\begin{tabular}{|c|c|c|}
  \hline
  $m^2L^2$ &  $(h,\bar{h})$ & $SU(3)\times SU(3)$  \\ \hline
  $3$& $\left(\frac{3}{2},\frac{3}{2}\right)$ & $2\times (\mathbf{1},\mathbf{1})$ \\
  \hline
  $-1$& $\left(\frac{1}{2},\frac{1}{2}\right)$ & $(\mathbf{3},\bar{\mathbf{3}})+(\bar{\mathbf{3}},\mathbf{3})$ \\
  \hline
\end{tabular}
\end{center}

\subsubsection{$SU(2,2)\times SU(2,2)$ gauging}
In this case, the sixteen scalars described by the coset
$\frac{SU(2,2)}{SU(2)\times SU(2)\times U(1)}\times
\frac{SU(2,2)}{SU(2)\times SU(2)\times U(1)}$ in which the two
$SU(2,2)\subset SU(4,4)$ are different from those appearing in the
gauge group can be parametrized as
\begin{eqnarray}
L&=& e^{a_1q_1}e^{a_2q_2}e^{a_3q_3}e^{a_4Q_1}e^{a_5Q_2}
e^{a_6Q_3}e^{\frac{i}{\sqrt{2}}b_1c_{16}}e^{\frac{i}{\sqrt{2}}b_2c_{27}}e^{a_7z_1}e^{a_8z_2}e^{a_9z_3}e^{a_{10}\tilde{z}_{1}}
e^{a_{11}\tilde{z}_{2}}\times
\nonumber\\
&
&e^{\frac{1}{\sqrt{2}}a_{12}\left[\tilde{z}_{1},\tilde{z}_{2}\right]}e^{\frac{i}{\sqrt{2}}b_3c_{40}}e^{\frac{i}{\sqrt{2}}b_4c_{55}},
\end{eqnarray}
where
\begin{equation}
\tilde{z}_{13}=\frac{1}{2}c_{61},\qquad
\tilde{z}_{14}=\frac{1}{2}c_{62},
\end{equation}
and $q_i$'s, $z_i$'s and $Q_i$'s are given in \eqref{qi_def},
\eqref{zi_def} and \eqref{Qi_def}, respectively. The potential is
given by
\begin{eqnarray}
V&=&-g^2 \left[8-2 \cosh(2 \sqrt{2} b_1)-2 \cosh(2 \sqrt{2} b_2)-2
\cosh(2 \sqrt{2} b_3)-2 \cosh(2
\sqrt{2} b_4)\right.\nonumber\\
&&\left.+4 \left(\cosh(\sqrt{2} b_1)+\cosh(\sqrt{2}
b_2)-\cosh(\sqrt{2} b_3)-\cosh(\sqrt{2} b_4)\right)^2\right].
\end{eqnarray}
There are two critical points:
\begin{itemize}
\item at $b_1=b_2=b_3=b_4=0$ with $V_0=0$ and $N=6$ supersymmetry
\item at $b_1=b_2=\frac{1}{\sqrt{2}}\cosh^{-1}{2}$, $b_3=b_4=0$ with
$V_0=8g^2$ which is equivalent to the critical point at
$b_3=b_4=\frac{1}{\sqrt{2}}\cosh^{-1}{2}$, $b_1=b_2=0$ with the same
$V_0$. In both cases, the residual symmetry is
$SU(2)_{\textrm{diag}}\times SU(2)\times SU(2)\times U(1)$. The
$SU(2)_{\textrm{diag}}$ is a diagonal subgroup of the $SU(2)$
factors from each $SU(2,2)$.
\end{itemize}
The scalar mass spectrum at the trivial critical point is given in
the table below.
\begin{center}
\begin{tabular}{|c|c|}
  \hline
  $m^2$ &   $SU(2)\times SU(2)\times SU(2)\times SU(2)$  \\ \hline
   $16g^2$&  $2\times (\mathbf{2},\mathbf{1},\mathbf{2},\mathbf{1})+2\times (\mathbf{1},\mathbf{2},\mathbf{1},\mathbf{2})$ \\
  \hline
\end{tabular}
\end{center}

\section{RG flow solutions}\label{flow}
In this section, we study RG flow solutions interpolating between
some AdS vacua given in section \ref{compact}. We consider only in
the $k=4$ case with gauge groups $SO(6)\times SU(4)\times U(1)$ and
$SO(4)\times SO(2)\times SU(4)\times U(1)$. The two cases have
different scalar potentials. According to what we have discussed in
section \ref{compact}, these are the only two independent potential
forms. So, in a sense, the study given in this section is enough
since the other cases can be studied in a very similar way.
According to the AdS/CFT correspondence, AdS critical points are
identified with two dimensional CFT's at the boundary of $AdS_3$ or
conformal fixed points of the dual two dimensional field theory. In
an RG flow, a UV CFT is perturbed by turning on some operators, or
the operators acquire vev's, which break conformal symmetry. If
there exist another conformal fixed point in the IR, the flow will
drive the UV CFT to another CFT in the IR. The central charge of the
corresponding CFT can be found by a well-known result
\begin{equation}
c=\frac{3L}{2G_N}\sim \frac{1}{\sqrt{-V_0}}
\end{equation}
where we have used the fact that in our case, the $AdS_3$ radius $L=
\frac{1}{\sqrt{-V_0}}$.
\\
\indent Because the two non-trivial critical points are
supersymmetric, the flow solution can be found by solving BPS
equations coming from setting the supersymmetry variation of
$\psi^I_\mu$ and $\chi^{iI}$ to zero. This solution will describe a
supersymmetric RG flow in the dual field theory.
\\
\indent Supersymmetry transformations of $\psi^I_\mu$ and
$\chi^{iI}$ are given  by \cite{dewit}
\begin{eqnarray}
\delta\psi^I_\mu
&=&\mathcal{D}_\mu\epsilon^I+gA_1^{IJ}\gamma_\mu\epsilon^J,\nonumber\\
\delta\chi^{iI}&=&
\frac{1}{2}(\delta^{IJ}\mathbf{1}-f^{IJ})^i_{\phantom{a}j}{\mathcal{D}{\!\!\!\!/}}\phi^j\epsilon^J
-gNA_2^{JIi}\epsilon^J\label{susyvar}
\end{eqnarray}
where $D_\mu
\epsilon^I=\left(\pd_\mu+\frac{1}{2}\omega_\mu^a\gamma_a+\ldots\right)\epsilon^I$.
\\
\indent As in usual studies of holographic RG flows, the metric
ansatz is taken to be
\begin{equation}
ds^2=e^{2A(r)}dx^2_{1,1}+dr^2,
\end{equation}
and the vector fields are set to zero. Recall the result of section
\ref{compact}, we find that the RG flow solution will interpolate
between two critical points with $(b_1,b_2)=(0,0)$ and
$(b_1,b_2)=(b,b)$ in which the two scalar fields $b_1$ and $b_2$
take the same value at both critical points. So, we use the ansatz
for the coset representative of the form
\begin{equation}
L=e^{b(r)Y_3}e^{b(r)Y_{15}}=e^{b(r)(Y_3+Y_{15})}
\end{equation}
where all the $Y$'s generators are given in the appendix. Since
$\left[Y_3,Y_{15}\right]=0$, the second identity in the above
parametrization follows immediately. In the $SO(6)\times SU(4)\times
U(1) $, the scalar $b(r)$ is one of the two singlets, with the
second one given by $Y_6+Y_{16}$, under the residual symmetry
$SO(4)\times SU(2)\times U(1)\times U(1)$. We have verified that the
truncation in which only the scalar corresponding to $Y_3+Y_{15}$ is
kept is consistent. The scalar which is truncated out appears at
least quadratically in the action, therefore, setting this scalar to
zero satisfies its equation of motion. Similarly, in the
$SO(4)\times SO(2)\times SU(4)\times U(1)$ gauge group, there are
four singlets under the residual symmetry $SU(2)\times U(1)^4$ given
by
\begin{displaymath}
Y_3+Y_{15},\qquad Y_3-Y_{15},\qquad Y_4+Y_{16},\qquad Y_4-Y_{16}\, .
\end{displaymath}
It can also be verified that the above truncation is a consistent
one.
\\
\indent It is now straightforward to use this information to compute
the BPS equations $\delta \psi^I_\mu=0$ and $\delta \chi^{iI}=0$. In
writing BPS equations below, we need to impose a projection
condition $\gamma_r\epsilon^I=\epsilon^I$ so that the flow solution
preserves half of the supersymmetries.
\subsection{RG Flows in $SO(6)\times SU(4)\times U(1)$ gauging}
In this case, the flow interpolates between $(6,0)$ critical point
and $(4,0)$ or $(2,0)$ points depending on the value of $\alpha$,
$\alpha<-1$ or $\alpha>-1$, respectively. The UV point is identified
with the $(6,0)$ point with $SO(6)\times SU(4)\times U(1)$ symmetry
while the IR point is identify with $(4,0)$ or $(2,0)$ points with
$SU(2)^3\times U(1)^2$ symmetry.
\subsubsection{An RG flow between $(6,0)$ and $(4,0)$ critical points}
The flow equation coming from $ \delta\chi^{iI}=0$ is given by
\begin{equation}
\frac{db(r)}{dr}=\sqrt{2}g\left[1-\alpha+(1+\alpha)\cosh(\sqrt{2}b(r))\right]\sinh(\sqrt{2}b(r)).
\end{equation}
We also recall that in this flow, $\alpha<-1$. This equation can be
solved for $r$ as a function of $b$. The solution is given by
\begin{eqnarray}
r&=&\frac{1}{4g\alpha}\ln
\left[\cosh\frac{b}{\sqrt{2}}\right]-\frac{1+\alpha}{8g\alpha}\ln
\left[1-\alpha+(1+\alpha)\cosh(\sqrt{2}b)\right]\nonumber
\\
& &+\frac{1}{4g}\ln \left[\sinh\frac{b}{\sqrt{2}}\right].
\end{eqnarray}
We have neglected the additive constant in the $r$ solution because
this can be removed by shifting the coordinate $r$.
\\
\indent As $b=0$, we find
\begin{equation}
b(r)\sim e^{4gr}\sim e^{-\frac{r}{2L_{\textrm{UV}}}}, \qquad
L_{\textrm{UV}}=\frac{1}{8|g|}\, .
\end{equation}
To identify this with the UV point in which $r\rightarrow \infty$,
we have chosen $g<0$. This asymptotic behavior of the scalar gives
information about the dimension of the operator that drives the
flow. The discussion on this point can be found in many references,
see for example \cite{freedman}. In this case, we find that the flow
is driven by a relevant operator of dimension $\Delta=\frac{3}{2}$.
\\
\indent At the IR point $b=
\frac{1}{\sqrt{2}}\cosh^{-1}\frac{\alpha-1}{\alpha+1}$, we find
\begin{equation}
b(r)\sim e^{-\frac{8g\alpha
r}{1+\alpha}}=e^{\frac{2\alpha}{(1+2\alpha)}\frac{r}{L_{\textrm{IR}}}},\qquad
L_{\textrm{IR}}=\frac{1+\alpha}{4|g|(1+2\alpha)}\, .
\end{equation}
In the IR, the operator becomes irrelevant and has dimension
$\Delta=\frac{2(1+3\alpha)}{1+2\alpha}$ which is greater than two
for $\alpha<-1$. The ratio of the central charges is given by
\begin{equation}
\frac{c_{\textrm{UV}}}{c_{\textrm{IR}}}=\frac{1+2\alpha}{2(1+\alpha)}>1,
\qquad \textrm{for} \,\,\, \alpha<-1\, .
\end{equation}
\indent The next thing is to determine the function $A(r)$ in the
metric. This can be done by using $\delta\psi^I_\mu=0$ equation for
$\mu=0,1$. The resulting equation is given by
\begin{eqnarray}
\frac{dA(r)}{dr}&=&-g\left[5+\alpha-(\alpha-1)\cosh(\sqrt{2}b(r))\right
.\nonumber
\\
& &\left .+\cosh(\sqrt{2}b(r))\left\{1-\alpha+(1+\alpha)
\cosh(\sqrt{2}b(r))\right\}\right].
\end{eqnarray}
Using $b(r)$ as an independent variable, we can write the above
equation as
\begin{equation}
\frac{dA(b)}{db}=-\frac{5+\alpha-(\alpha-1)\cosh(\sqrt{2}b)+\cosh(\sqrt{2}b)(1-\alpha+(1+\alpha)\cosh(\sqrt{2}b))}
{\sqrt{2}\left[1-\alpha+(1+\alpha)\cosh(\sqrt{2}b)\right]\sinh(\sqrt{2}b)}\,
.
\end{equation}
The solution is easily obtained to be
\begin{eqnarray}
A&=&\frac{1+2\alpha}{2\alpha}\ln
\left[1-\alpha+(1+\alpha)\cosh(\sqrt{2}b)\right]-2\ln
\left[2\sinh\frac{b}{\sqrt{2}}\right]\nonumber
\\
& &-\frac{1+\alpha}{\alpha}\ln \left[2\cosh
\frac{b}{\sqrt{2}}\right].
\end{eqnarray}
We have again neglected the additive constant by rescaling the
coordinates $x^\mu$ for $\mu=0,1$.
\\
\indent We can readily check from the kinetic term that the scalar
$\sqrt{2}b(r)$ is canonically normalized, so we can read off the
value of its mass squared from the scalar potential. To quadratic
order near the UV point, the potential \eqref{potential_k2_SO6} is
given by
\begin{equation}
V=-64g^2-48g^2b^2
\end{equation}
which gives $m^2L_{\textrm{UV}}^2=-\frac{3}{4}$. We find that the
relation $m^2L^2=\Delta(\Delta-2)$ precisely gives
$\Delta=\frac{3}{2}$ consistent with what we have found before from
the asymptotic behavior of the scalar $b(r)$.
\\
\indent At the IR point, we obtain the potential to second order
\begin{equation}
V=-\frac{16g^2(1+2\alpha)^2}{(1+\alpha)^2}+\frac{64g^2\alpha(1+3\alpha)}{(1+\alpha)^2}(b-b_0)^2
\end{equation}
where $b_0=\frac{1}{\sqrt{2}}\cosh^{-1}\frac{\alpha-1}{\alpha+1}$.
We find
$m^2L^2_{\textrm{IR}}=\frac{4\alpha(1+3\alpha)}{(1+2\alpha)^2}$ or
$\Delta=\frac{2(1+3\alpha)}{1+2\alpha}$.
\subsubsection{An RG flow between $(6,0)$ and $(2,0)$ critical points}
We now consider an RG flow between $(6,0)$ and $(2,0)$ critical
points similar to what we have done in the previous case. In this
case, $\alpha>-1$.
\\
\indent We begin with the flow equation obtained from
$\delta\chi^{iI}=0$
\begin{equation}
\frac{db(r)}{dr}=-\sqrt{2}g\left[-3-\alpha+(1+\alpha)\cosh(\sqrt{2}b(r))\right]\sinh(\sqrt{2}b(r)).
\end{equation}
As before, we can solve for $r$ as a function of $b$ and find
\begin{eqnarray}
r&=&-\frac{1+\alpha}{8g(2+\alpha)}\ln
\left[(1+\alpha)\cosh(\sqrt{2}b)-3-\alpha\right]+\frac{1}{4g}\ln
\left[\sinh\frac{b}{\sqrt{2}}\right]\nonumber
\\
& &-\frac{1}{4g(2+\alpha)}\ln \left[\cosh\frac{b}{\sqrt{2}}\right].
\end{eqnarray}
\indent Near the UV point $b=0$, we find
\begin{equation}
b(r)\sim e^{4gr}=e^{-\frac{r}{2L_{\textrm{UV}}}},\qquad
L_{\textrm{UV}}=\frac{1}{8|g|}\, .
\end{equation}
We again choose $g<0$ to identify the UV point with $r\rightarrow
\infty$. The flow is driven by a relevant operator of dimension
$\Delta=\frac{3}{2}$.
\\
\indent Near the IR point
$b=\frac{1}{\sqrt{2}}\cosh^{-1}\frac{\alpha+3}{\alpha+1}$, the
scalar behaves as
\begin{equation}
b(r)\sim
e^{-\frac{8g(2+\alpha)r}{1+\alpha}}=e^{\frac{2(2+\alpha)r}{(3+2\alpha)L_{\textrm{IR}}}},\qquad
L_{\textrm{IR}}=\frac{1+\alpha}{4|g|(3+2\alpha)}\, .
\end{equation}
The operator has dimension
$\Delta=\frac{2(5+3\alpha)}{3+2\alpha}>2$, for $\alpha>-1$, in the
IR.
\\
\indent The equation for $A(r)$ obtained from $\delta \psi^I_\mu=0$
is given by
\begin{eqnarray}
\frac{dA(r)}{dr}&=&g\left[\alpha-3-(3+\alpha)\cosh(\sqrt{2}b(r))\right
. \nonumber
\\
& &\left
.+\cosh(\sqrt{2}b(r))\{-3-\alpha+(1+\alpha)\cosh(\sqrt{2}b(r))\}\right]
\end{eqnarray}
or, in term of variable $b$,
\begin{equation}
\frac{dA}{db}=-\frac{\alpha-3-(3+\alpha)\cosh(\sqrt{2}b)+\cosh(\sqrt{2}b)\left[-\alpha-3+(1+\alpha)\cosh(\sqrt{2}b)\right]}
{\sqrt{2}\left[-3-\alpha+(1+\alpha)\cosh(\sqrt{2}b)\right]\sinh(\sqrt{2}b)}\,
.
\end{equation}
The solution for $A(b)$ is found to be
\begin{eqnarray}
A&=&-\frac{1+\alpha}{2+\alpha}\ln\left[2\cosh\frac{b}{\sqrt{2}}\right]+\frac{3+2\alpha}{2(2+\alpha)}\ln
\left[-3-\alpha+(1+\alpha)\cosh(\sqrt{2}b)\right]\nonumber \\
& &-2\ln \left[2\sinh\frac{b}{\sqrt{2}}\right].
\end{eqnarray}
The ratio of the central charges is given by
\begin{equation}
\frac{c_{\textrm{UV}}}{c_{\textrm{IR}}}=\frac{3+2\alpha}{2(1+\alpha)}>1,\qquad
\textrm{for}\qquad \alpha>-1\, .
\end{equation}
\indent We can compute the value of mass squared at both end points
and find that the dimension of the operator found previously agrees
with the result obtained from the relation
$m^2L^2=\Delta(\Delta-2)$. We will not give a repetition here.
\subsection{RG Flows in $SO(4)\times SO(2)\times SU(4)\times U(1)$ gauging}
In this case, there are three critical points involving in the
flows. The trivial critical point at $b_1=b_2=0$ has $(4,2)$
supersymmetry with $SO(4)\times SO(2)\times SU(4)\times U(1)$ gauge
symmetry. The two non-trivial critical points are given by
$b_1=b_2=\frac{1}{\sqrt{2}}\cosh^{-1}\frac{2+\alpha}{\alpha}$ with
$(2,2)$ supersymmetry and
$b_1=b_2=\frac{1}{\sqrt{2}}\cosh^{-1}\frac{\alpha-2}{\alpha}$ with
$(2,0)$ supersymmetry. Both of them have unbroken gauge symmetry
$U(1)^4\times SU(2)$.
\\
\indent All procedures involved are the same as in the previous
subsection, so we will not go into much detail but simply give key
results in each step. Similar to the previous gauge group, there are
two possible flows namely the flow between $(4,2)$ and $(2,2)$
points and between $(4,2)$ and $(2,0)$ points.
\subsubsection{An RG flow between $(4,2)$ and $(2,2)$ critical points}
In this case, the flow interpolates between two critical points with
chiral and non-chiral supersymmetries. The $(2,2)$ critical point
requires $\alpha>0$. The flow equation from $\delta \chi^{iI}=0$
reads
\begin{equation}
\frac{db(r)}{dr}=-\sqrt{2}g\left[2+\alpha-\alpha\cosh(\sqrt{2}b(r))\right]\sinh(\sqrt{2}b(r)).
\end{equation}
The solution for $r$ in term of $b$ is found to be
\begin{eqnarray}
r&=&\frac{1}{4g(1+\alpha)}\ln
\left[\cosh\frac{b}{\sqrt{2}}\right]+\frac{\alpha}{8g(1+\alpha)}\ln
\left[\alpha\cosh(\sqrt{2}b)-\alpha-2\right]\nonumber
\\
& &-\frac{1}{4g}\ln\left[\sinh\frac{b}{\sqrt{2}}\right].
\end{eqnarray}
\indent Choosing $g>0$, the UV point is described by $r\rightarrow
\infty$. Near this point, the scalar behaves
\begin{equation}
b(r)\sim e^{-4gr}=e^{-\frac{r}{2L_{\textrm{UV}}}},\qquad
L_{\textrm{UV}}=\frac{1}{8g}\, .
\end{equation}
The flow is seen to be driven by a relevant operator of dimension
$\frac{3}{2}$.
\\
\indent At the IR point, $r\rightarrow -\infty$, we find
\begin{equation}
b(r)\sim
e^{\frac{8g(1+\alpha)r}{\alpha}}=e^{\frac{2(1+\alpha)r}{(1+2\alpha)L_{\textrm{IR}}}},\qquad
L_{\textrm{IR}}=\frac{\alpha}{4g(1+2\alpha)}\, .
\end{equation}
The dimension of the operator in the IR is
$\Delta=\frac{2(2+3\alpha)}{1+2\alpha}$ which is greater than two
for $\alpha>0$.
\\
\indent We then move to the $\delta\psi^I_\mu=0$ equation which
gives
\begin{equation}
\frac{dA}{dr}=-\frac{1}{2}g\left[8-3\alpha+4(2+\alpha)\cosh(\sqrt{2}b(r))-\alpha\cosh(2\sqrt{2}b(r))\right]
\end{equation}
or in term of variable $b$
\begin{equation}
\frac{dA(b)}{db}=-\frac{8-3\alpha+4(2+\alpha)\cosh(\sqrt{2}b(r))-\alpha\cosh(2\sqrt{2}b(r))}
{2\sqrt{2}\left[2+\alpha-\alpha\cosh(\sqrt{2}b)\right]\sinh(\sqrt{2}b)}\,
.
\end{equation}
The solution is given by
\begin{eqnarray}
A&=&\frac{\alpha}{1+\alpha}\ln\left(2\cosh\frac{b}{\sqrt{2}}\right)-\frac{1+2\alpha}{2(1+\alpha)}
\ln\left[\alpha\cosh(\sqrt{2}b)-\alpha-2\right]\nonumber \\
& &+2\ln \left(2\sinh\frac{b}{\sqrt{2}}\right).
\end{eqnarray}
It is easily seen that as $b\rightarrow 0$, $A\rightarrow \infty$
and as $b\rightarrow
\frac{1}{\sqrt{2}}\cosh\frac{\alpha+2}{\alpha}$, $A\rightarrow
-\infty$. The ratio of the central charges is given by
\begin{equation}
\frac{c_{\textrm{UV}}}{c_{\textrm{IR}}}=\frac{1+2\alpha}{2\alpha}>1,\qquad
\textrm{for}\qquad \alpha>0\, .
\end{equation}
\subsubsection{An RG flow between $(4,2)$ and $(2,0)$ critical points}
We now give our last RG flow solution. The flow equation reads
\begin{equation}
\frac{db(r)}{dr}=-\sqrt{2}g\left[\alpha-2-\alpha\cosh(\sqrt{2}b(r))\right]\sinh(\sqrt{2}b(r))
\end{equation}
whose solution for $r(b)$ is given by
\begin{eqnarray}
r&=&\frac{1}{4g(\alpha-1)}\ln
\left(\cosh\frac{b}{\sqrt{2}}\right)-\frac{\alpha}{8g(\alpha-1)}\ln\left[2-\alpha+\alpha\cosh(\sqrt{2}b)\right]\nonumber
\\
& &+\frac{1}{4g} \ln\left(\sinh\frac{b}{\sqrt{2}}\right).
\end{eqnarray}
Near $b=0$, the UV point, we find
\begin{equation}
b(r)\sim e^{4gr}=e^{-\frac{r}{2L_{\textrm{UV}}}},\qquad
L_{\textrm{UV}}=\frac{1}{8|g|}
\end{equation}
where we have chosen $g<0$ to identify this as the UV point when
$r\rightarrow \infty$.
\\
\indent Near the IR point,
$b=\frac{1}{\sqrt{2}}\cosh\frac{\alpha-2}{\alpha}$ and $r\rightarrow
-\infty$, we find
\begin{equation}
b(r)\sim
e^{-\frac{8g(\alpha-1)r}{\alpha}}=e^{\frac{(2\alpha-1)r}{(2\alpha-1)L_{\textrm{IR}}}},\qquad
L_{\textrm{IR}}=\frac{\alpha}{4|g|(2\alpha-1)}\, .
\end{equation}
So, the flow is again driven by a relevant operator of dimension
$\frac{3}{2}$, and the operator becomes irrelevant in the IR with
dimension $\Delta=\frac{2(2-3\alpha)}{1-2\alpha}>2$ for $\alpha<0$.
\\
\indent We finally determine the $A(r)$ function in this case. It is
the solution of the $\delta\psi^I_\mu=0$ equation given by
\begin{equation}
\frac{dA(r)}{dr}=-\frac{1}{2}g\left[8+3\alpha-4(\alpha-2)\cosh(\sqrt{2}b(r))+\alpha(2\sqrt{2}b(r))\right]
\end{equation}
or
\begin{equation}
\frac{dA(b)}{db}=\frac{8+3\alpha-4(\alpha-2)\cosh(\sqrt{2}b(r))+\alpha(2\sqrt{2}b(r))}
{2\sqrt{2}\left[\alpha-2-\alpha\cosh(\sqrt{2}b)\right]\sinh(\sqrt{2}b)}\,
.
\end{equation}
The solution is found to be
\begin{eqnarray}
A&=&\frac{\alpha}{\alpha-1}\ln\left[2\cosh\frac{b}{\sqrt{2}}\right]+\frac{1-2\alpha}{2(\alpha-1)}
\ln\left[2-\alpha+\alpha\cosh(\sqrt{2}b)\right]\nonumber
\\
& &-2\ln\left(2\sinh\frac{b}{\sqrt{2}}\right).
\end{eqnarray}
The ratio of the central charges is given by
\begin{equation}
\frac{c_{\textrm{UV}}}{c_{\textrm{IR}}}=\frac{2\alpha-1}{2\alpha}>1,\qquad
\textrm{for}\qquad \alpha<0\, .
\end{equation}
\indent The analysis of the mass squared from the expansion of the
scalar potential near the critical points can be done in the same
way as in the previous subsection. We will again not repeat this
computation here.
\section{Conclusions}
In this paper, we have extensively studied $N=6$ gauged supergravity
in three dimensions. We have used the Euler angle parametrization in
parametrizing the scalar coset manifold $\frac{SU(4,k)}{S(U(4)\times
U(k))}$ and submanifolds thereof. This parametrization turns out to
be very useful. We have identified admissible gauge groups of
non-compact type, studied their scalar potentials and found some of
the critical points associated to each scalar potential.
\\
\indent In compact gauge groups identified in \cite{dewit}, we have
found a number of supersymmetric AdS vacua. It is possible in this
case to discuss holographic RG flow solutions in the context of the
AdS/CFT correspondence. We have studied four analytic RG flow
solutions in two gauge groups, $SO(6)\times SU(4)\times U(1)$ and
$SO(4)\times SO(2)\times SU(4)\times U(1)$, in the $k=4$ case. Given
the structure of the critical points including the form of the
potential, this study is, in a sense, sufficient because among
$k=2,3,4$ cases, there are actually only two different potential
forms. The flows in other gauge groups or in different values of $k$
can be obtained similarly. The resulting solutions involve one
active scalar and can be solved analytically. The flows are operator
flows driven by a relevant operator of dimension $\frac{3}{2}$ and
respect the c-theorem. Notice that the solutions look very much like
the solutions found in \cite{bs} and \cite{AP}.
\\
\indent In non-compact gauge groups, it is remarkable that apart
from $L=\mathbf{I}$, we have not found any other (non-trivial) AdS
critical points whether supersymmetric or not. We strongly believe
that there is no non-trivial AdS critical point in the non-compact
groups studied here. Therefore, there are no possibilities of RG
flows in the AdS/CFT sense. However, it is interesting to study
domain wall solutions connecting between dS vacua or even between dS
and Minkowski vacua as discussed in \cite{Cvetic}. A domain wall
between dS vacua could also be interpreted as an RG flow in the
context of the dS/CFT correspondence \cite{dSCFT}. An example of
this study can be found in \cite{dSRG}.
\\
\indent Recently, the new approach for finding critical points of
gauged supergravities has been introduced in \cite{new_approach}.
This technique is based on the variation of the embedding tensor
rather than the extremization of the scalar potential as done in
this and many other works. New critical points of $N=8$ gaued
supergravity in four dimensions have been found within this
framework. It is naturally interesting to investigate critical
points of the $N=6$ theory studied here particularly for compact
gauge groups, which have not been explored in full details due to
the complication of the computation, with this new approach as well
as to reexamine the scalar potentials of the theories with different
values of $N$ studied in \cite{bs}, \cite{gkn}, \cite{AP},
\cite{AP2} and \cite{N16Vacua}. We hopefully expect to find some new
vacua of these theories, too. We leave these issues for future
works.
\acknowledgments This work is partially supported by Faculty
of Science, Chulalongkorn University through the A1B1 project and
Thailand Center of Excellence in Physics through the
ThEP/CU/2-RE3/11 project. The second author would like to thank
Chulalongkorn University for the support via Ratchadapisek Sompote
Endowment Fund in carrying out this research project.
\appendix
\section{On the Euler angle parametrization}
In this appendix, we review the Euler angle parametrization of a Lie
group $G$ using Euler angles of its subgroup $H$. We begin with some
general idea of this parametrization based on the the result of
\cite{exceptional coset} and then specify to our case for the coset
of the form $\frac{SU(n,m)}{SU(n)\times SU(m)\times U(1)}$. The same
procedure can be applied to all cases discussed in this paper. So,
we will give only one example namely the coset
$\frac{SU(4,3)}{SU(4)\times SU(3)\times U(1)}$ in the case of $k=4$.
\\
\indent For a Lie group $G$, the $H$ Euler angle parametrization of
$G$ is given by \cite{exceptional coset}
\begin{equation}
G=Be^VH\, .
\end{equation}
$H$ is the parametrization of the subgroup $H\subset G$ which can in
turn be parametrized by Euler angles of some subgroup $H_1\subset
H$. However, we will explain only the parametrization of $G$ itself
since our main aim here is to demonstrate the parametrization
procedure. In our application with non-compact $G$, we would like to
parametrize the coset space $G/H$ which can be obtained from the
above parametrization by a quotient with $H$. Therefore, in the
following, we will choose $H$ to be the maximal compact subgroup of
$G$. Following \cite{exceptional coset}, we denote the algebras of
$G$ and $H$ by $\mathfrak{g}$ and $\mathfrak{h}$, respectively. The
$G$ generators are then decomposed into $H$ generators and
non-compact generators which constitute a subspace
$\mathfrak{p}\subset \mathfrak{g}$. The subspace $V\subset
\mathfrak{p}$ consists of the minimal set of non-compact generators
such that the whole subspace $\mathfrak{p}$ can be generated by the
adjoint action of $H$ on $V$.
\\
\indent The factor $B=H/H_0$ where $H_0$ is the redundancy of the
parametrization. Obviously, we find $\textrm{dim}\,
G=\textrm{dim}\,H/H_0+\textrm{dim}\,V+\textrm{dim}\,H$. $H_0$
consists of the automorphisms of $V$. They are not needed to
generate the whole $\mathfrak{p}$ from $V$, or equivalently, they
commute with $V$. We will see how this procedure works in the
following example.
\\
\indent We recall that the $SU(8)$ generators can be constructed
from the generalized Gell-Mann matrices $c_i$, $i=1,\ldots, 63$. Its
non-compact form $SU(4,4)$ is obtained by multiplying each
non-compact generator by a factor of $i$. Explicitly, the
non-compact generators for the $k=4$ case are given by
\begin{equation}
Y_A=\left \{ \begin{array}{c}
               \frac{1}{\sqrt{2}}c_{A+15},\qquad A=1,\ldots, 8 \\
               \frac{1}{\sqrt{2}}c_{A+16},\qquad A=9,\ldots, 16 \\
               \frac{1}{\sqrt{2}}c_{A+19},\qquad A=17,\ldots, 24 \\
               \frac{1}{\sqrt{2}}c_{A+24},\qquad A=25,\ldots, 32
             \end{array}
\right .\, .
\end{equation}
\indent We are now in a position to parametrize the coset
$K=\frac{SU(4,3)}{SU(4)\times SU(3)\times U(1)}$ in the case of
$SU(1,4)\times SU(3)\times U(1)$ gauging. The subgroup
$H=SU(4)\times SU(3)\times U(1)$ and contains 24 parameters. There
are 24 scalars in the coset $K$. These scalars correspond to the
non-compact generators $Y_i$ for $i=1,\ldots, 6$, $9,\ldots, 14$,
$17,\ldots, 22$, $25,\ldots, 30$. The subspace $V$ consists of three
generators which can be chosen to be $Y_1, Y_{11}, Y_{21}$. It is
now easy to verify that a subgroup of $H$ that commutes with these
three generators is $U(1)\times U(1)\times U(1)$. So, the redundancy
in the parametrization is given by $H_0=U(1)^3$. We then find
$B=\frac{SU(4)\times SU(3)\times U(1)}{U(1)\times U(1)\times U(1)}$.
We can identify one of the $U(1)$ in $U(1)^3$ with the $U(1)$ factor
in $H$. Furthermore, we also choose to remove the remaining $U(1)^2$
in $H_0$ by moding out one $U(1)$ factor from $SU(4)$ and the other
one from $SU(3)$. We are now left with $B=\frac{SU(4)}{U(1)}\times
\frac{SU(3)}{U(1)}$. There are other ways of removing the redundancy
other than that given here, but they are equivalent after redefining
the scalars.
\\
\indent With all these, the coset representative for the coset $K$
is given by
\begin{eqnarray}
L&=& e^{a_1c_3}e^{a_2c_2} e^{a_3 c_3}e^{a_4 c_5}
e^{\frac{1}{\sqrt{3}}a_5c_8}e^{a_6c_3}
e^{a_7c_3}e^{a_8j_3}e^{a_9j_2}
e^{a_{10} j_3 } e^{a_{11} j_5 }e^{\frac{1}{\sqrt{3}}a_{12} j_8}e^{a_{13} j_{10} }\times \nonumber\\
&&e^{a_{14} j_3 }e^{a_{14} j_3 }e^{a_{15} j_2 }e^{a_{16} j_3
}e^{a_{17} j_5 }e^{\frac{1}{\sqrt{3}}a_{18} j_8}e^{a_{19} j_3
}e^{a_{20} j_2 }e^{a_{21} j_3
}e^{b_1Y_{1}}e^{b_2Y_{11}}e^{b_3Y_{21}},
\end{eqnarray}
where the $SU(4)$ generators $j_i$'s are defined in \eqref{ji_def}.
The generators denoted by $c_i$'s generate the $SU(3)$. The explicit
parametrization of both $SU(4)$ and $SU(3)$ can be found in
\cite{SUN_Euler}.


\begin{thebibliography}{99}
\bibitem{gaugeSUGRA_flux} Henning Samtleben, `` Lectures on Gauged Supergravity and Flux
Compactifications'', Class. Quant. Grav. (2008) \textbf{25} 214002,
arXiv: 0808.4076.
\bibitem{maldacena} J. M. Maldacena, ``The large $N$ limit of
superconformal field theories and supergravity'', Adv. Theor. Math.
Phys. \textbf{2} (1998) 231-252, arXiv: hep-th/9711200.
\bibitem{henning_ADSCMT} E. O. Colgain and H. Samtleben, ``3D gauged supergravity from wrapped M5-branes with AdS/CMT
application'', JHEP 02 (2011) \textbf{031}, arXiv: 1012.2145.
\bibitem{dewit1} Bernard de Wit, A. K. Tollsten and H. Nicolai,
``Locally supersymmetric $D=3$ nonlinear sigma models'', Nucl. Phys.
\textbf{B392} (1993) 3-38, arXiv: hep-th/9208074.
\bibitem{dewit} Bernard de Wit, Ivan Herger and Henning Samtleben, ``Gauged Locally Supersymmetric $D=3$ Nonlinear Sigma
Models'', Nucl. Phys. \textbf{B671} (2003) 175-216, arXiv:
hep-th/0307006.
\bibitem{csym} H. Nicolai and H. Samtleben, ``Chern-Simons vs Yang-Mills gaugings in three dimensions'', Nucl. Phys. B \textbf{638} (2002) 207-219
, arXiv: hep-th/0303213.
\bibitem{3D_conformal_gauge} E. A. Bergshoeff, O. Hohm, D. Roest, H.
Samtleben and E. Sezgin, ``The Superconformal Gaugings in Three
Dimensions'', JHEP 09 (2008) \textbf{101}, arXiv: 0807.2841.
\bibitem{bs} M. Berg and H. Samtleben, ``An exact holographic RG Flow between 2d Conformal Field
Theories'', JHEP \textbf{05} (2002) 006, arXiv: hep-th/0112154.
\bibitem{gkn} Edi Gava, Parinya Karndumri and K. S. Narain, ``AdS$_3$ Vacua and RG Flows in Three Dimensional Gauged
Supergravities'', JHEP 04 (2010) \textbf{117}, arXiv: 1002.3760.
\bibitem{AP} Auttakit Chatrabhuti and Parinya Karndumri, ``Vacua and
RG flows in $N=9$ three dimensional gauged supergravity'', JHEP 10
(2010) \textbf{098}, arXiv: 1007.5438.
\bibitem{AP2} Auttakit Chatrabhuti and Parinya Karndumri, ``Vacua of $N=10$ three
dimensional gauged supergravity'', Class. Quantum Grav. \textbf{28}
(2011) 125027, arXiv: 1011.5355.
\bibitem{Fisch} T. Fischbacher, ``Some stationary points of gauged
$N=16$ $D=3$ supergravity'', Nucl.Phys. \textbf{B638} (2002)
207-219, arXiv: hep-th/0201030.
\bibitem{N16Vacua}T. Fischbacher, H. Nicolai and H. Samtleben,
``Vacua of Maximal Gauged $D=3$ Supergravities'', Class. Quant.
Grav. \textbf{19} (2002) 5297-5334, arXiv: hep-th/0207206.
\bibitem{deger} N. S. Deger, ``Renormalization group flows from $D = 3$, $N=2$
 matter coupled gauged supergravities'', JHEP \textbf{0211} (2002) 025. arXiv: hep-th/0209188.
\bibitem{stancu} Fl. Stancu, ``Group Theory in Subnuclear Physics'',
Oxford University Press, USA 1997.
\bibitem{SUN_Euler} S. Bertini, S. L. Cacciatori, B. L. Cerchiai,
``On the Euler angles for $SU(N)$'', J. Math. Phys. \textbf{47}
(2006) 043510, arXiv: math-ph/0510075.
\bibitem{exceptional coset} Sergio L. Cacciatori and B. L. Cerchiai,
``Exceptional groups, symmetric spaces and applications'', arXiv:
0906.0121.
\bibitem{nicolai2} H. Nicolai and H. Samtleben, ``Compact and noncompact gauged maximal
supergravities in three dimensions'', JHEP 04 (2001) \textbf{022},
arXiv: hep-th/0103032.
\bibitem{warner} N. P. Warner, ``Some New Extrema of the Scalar
Potential of Gauged $N=8$ Supergravity'', Phys. Lett. \textbf{B128}
(1983) 169.
\bibitem{SC} E. S. Fradkin and Y. Ya. Linetsky, ``Results of the
classification of superconformal algebras in two dimensions'', Phys.
Lett. \textbf{B282} (1992) 352-356, arXiv: hep-th/9203045.
\bibitem{freedman} Eric D'Hoker and Daniel Z. Freedman, ``Supersymmetric Gauge Theories and the AdS/CFT Correspondence'',
TASI 2001 Lecture Notes. arXiv: hep-th/0201253.
\bibitem{Cvetic} M. Cvetic and Jing Wang, ``Vacuum Domain Walls in D-dimensions: Local and Global Space-Time
Structure'', Phys. Rev. \textbf{D61} (2000) 124020, arXiv:
hep-th/9912187.
\bibitem{dSCFT} A. Strominger, ``The dS/CFT Correspondence'', JHEP
10 (2001) \textbf{034}, arXiv: hep-th/0106113.
\bibitem{dSRG} S. Nojiri and S. D. Odintsov, ``Asymptotically de Sitter dilatonic space-time, holographic RG flow and conformal anomaly from
(dilatonic) dS/CFT correspondence'', Phys. Lett. \textbf{B531}
(2002) 143-151, arXiv: hep-th/0201210.
\bibitem{new_approach} G. Dall' Agata and G. Inverso, ``On the vacua of N=8 gauged supergravity in 4
dimensions'', arXiv: 1112.3345.
\end{thebibliography}
\end{document}